\documentstyle[aps,prl,twocolumn,epsfig]{revtex}
\begin{document}
\draft
\title{Zero-field Time Correlation Functions of Four Classical Heisenberg Spins on a 
Ring }
\author{Richard A. Klemm$^{1,2,3}$ and Marshall Luban$^2$}
\address{$^1$Max-Planck-Institut f{\"u}r Physik komplexer Systeme, N{\"o}thnitzer
 Stra{\ss}e 38, D-01187 Dresden, Germany}
\address{$^2$Ames Laboratory and Department of Physics and Astronomy, 
Iowa State 
University, Ames, IA 50011 USA}
\address{$^3$Materials Science Division, Argonne
National Laboratory, Argonne, IL
60439 USA}

\date{\today}
\maketitle
\begin{abstract}
A model relevant for the study of certain molecular magnets is the  
ring of
$N=4$ classical spins with equal near-neighbor isotropic Heisenberg exchange 
interactions.  Assuming classical Heisenberg spin dynamics, we solve 
explicitly for the time evolution of each of the spins.
 Exact triple integral representations are
 derived for the auto, near-neighbor, and next-nearest-neighbor time
 correlation functions for any temperature.     At infinite temperature,
 the correlation functions are reduced to quadrature.  We then evaluate the Fourier
 transforms of these functions in closed form, which are double integrals.
 At low temperatures, the Fourier transform functions explicitly demonstrate
 the presence of magnons.   Our exact results for the infinite temperature correlation
functions in the long-time asymptotic limit differ qualitatively from those 
obtained assuming diffusive spin dynamics.  Whether such explicitly
non-hydrodynamic behavior would be maintained for large-$N$ rings is discussed. 

\end{abstract}
\vskip0pt
\vskip0pt\vskip0pt
\section{Introduction}
Recently, there has been a rapidly growing interest in the physics of molecular magnets.
 \cite{Gatteschi,Friedman}  These compounds can be synthesized as single
 crystals of identical molecular units, each containing several paramagnetic
 ions that mutually interact via Heisenberg exchange.  The intermolecular
 (dipole-dipole) magnetic interactions are in the great majority of cases
 utterly negligible as compared to intramolecular magnetic interactions.  Measurements
 of the magnetic properties therefore reflect those of the common, individual
 molecular units of nanometer size.  Their dynamics can be studied by inelastic
 neutron scattering, as well as by nuclear magnetic resonance and electron
 paramagnetic resonance experiments.   Some of these molecular magnets are made
 of very small clusters of magnetic ions.  The smallest clusters are dimers of
 V$^{4+}$ ($S=1/2$) and of Fe$^{3+}$ ($S=5/2$), \cite{V2,Fe2} a nearly
 equilateral triangle array of V$^{4+}$ spins, \cite{V3}, a nearly square
 array of Nd$^{3+}$ (total spin $j=9/2$), \cite{Nd4} a regular tetrahedron of Cr$^{3+}$
 ($S=3/2$), \cite{Cr4} a frustrated tetrahedral pyrochlore of Tb$^{3+}$
 ($S=5/2$), \cite{Tb4} and a ``squashed'' tetrahedron of Fe$^{3+}$
 spins. \cite{Fe4,Fe4e}  There has also been an example of a four-spin ring
 which is coupled to nearby rings, although the spin value ($S=1/2$) is
 small, and thus requires a quantum treatment. \cite{CaV4O9}  In addition,
 larger rings, most notably with 6, 8, or 10 Fe$^{3+}$ spins, 
 have been studied. \cite{Fe6,Fe8,Fe10} 

In some of these systems, the spin value of an individual magnetic ion is large
 enough that the dynamics can be closely approximated by the classical theory,
 as long as one does not go to temperatures that are too low. 
 Thus, it is useful to study such systems theoretically, in order to investigate the types
 of dynamical spin behavior that can occur.
Such investigations can provide helpful physical insight, as well as some
 guidance for systems that might be studied experimentally.  It will also be
 interesting to compare the classical results with those emerging from studies
 of their quantum analogues, such as has been done for the dimer and the
 equilateral triangle. \cite{2spin,Mentrup}

Perhaps more interesting, however, is the question as to whether the long-time
asymptotic behavior of the two-spin correlation functions at infinite
temperature will be consistent
with the results of a hydrodynamic-like theory, in which the exact equations
governing the spin dynamics are approximated by linear diffusion-like equations [see
Eq. ({\ref{diffuse})]. This question has been the subject of much debate
in the literature, \cite{Mueller,Landau,Mueller1,Bonfim}, and a
solution of the spin dynamics for the four-spin ring might aid in our
understanding of this more fundamental problem.  Here we derive exact results
which explicitly demonstrate that the infinite temperature, long-time
asymptotic limits of the $N=4$ two-spin correlation functions are non-hydrodynamic.

The layout of the paper is as follows.  In Sec. II, we give the notation,
partition function, and derive the exact time evolution of the individual spin
vectors.  In Sec. III, we give the results for the time correlation
functions.  At infinite temperature, these results can be expressed as single
integrals, but at finite  temperatures, they are triple integrals.  We also
present our derivation of the Fourier transforms of the deviations of the
correlation functions from their infinite time asymptotic limits.  Finally, we
invite the reader to read our discussion and conclusions in Sec. IV, even if
one has only a minimal interest in the mathematical developments presented in
Secs. II and III.  In this final section, we also discuss the non-hydrodynamic
aspects of our exact results for the infinite temperature, long-time
asymptotic behaviors of the two-spin correlation functions, and raise the
question as to whether such non-hydrodynamic features might be maintained for
larger rings. 

\section{Spin Dynamics}

\subsection{Notation and partition function}

We study the dynamics of four interacting spins on a ring.  Each spin has  unit magnitude and
can assume an  arbitrary direction.  We suppose that it interacts only with its two nearest neighbors.  We label these
 spins ${\bf S}_1, {\bf S}_2,$ etc., where $|{\bf S}_i|=1$, illustrated in Fig. 1.  

\begin{figure}
\vspace{0.3cm}
\epsfxsize=6.5cm
\centerline{\hspace{1cm}\epsffile{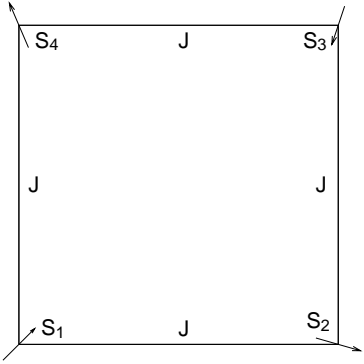}}
\vspace{1.0cm}
\caption{Sketch of a ring with four classical spins 
at the corners, each interacting 
with its nearest neighbors with strength $J$.}
\label{fig1}  
\end{figure}

 The Hamiltonian for this system is thus

\begin{eqnarray}
H& = &-J\sum_{i=1}^4{\bf S}_i\cdot{\bf S}_{i+1},\\
& = &-(J/2)(S^2-S_{13}^2-S_{24}^2),\label{H}
\end{eqnarray}
where ${\bf S}_5={\bf S}_1$, ${\bf S}_{13}={\bf S}_1+{\bf S}_3$, ${\bf S}_{24}={\bf S}_2+{\bf
 S}_4$, and ${\bf S}={\bf S}_{13}+{\bf S}_{24}$.\cite{Luban}

The partition function $Z$ can be readily found from Eq. (\ref{H}), leading to
\begin{eqnarray}
Z& = &{\rm Tr}\>\exp(-\beta H)\\
& = &\Bigl(\prod_{i=1}^4\int{{d\Omega_i}\over{4\pi}}\Bigr)\int d^3{\bf S}
\int d^3{\bf S}_{13}\int d^3{\bf S}_{24}\times\nonumber\\
& &\times\delta^{(3)}({\bf S}-{\bf S}_{13}-{\bf S}_{24})\delta^{(3)}
({\bf S}_{13}-{\bf S}_1-{\bf S}_3)\times\nonumber\\
& &\times\delta^{(3)}({\bf S}_{24}-{\bf S}_2-{\bf S}_4){\rm exp}(-\beta H)\label{ZS}\\
&=&\Bigl(\prod_{i=1}^4\int{{d\Omega_i}\over{4\pi}}\Bigr)\int d^3{\bf S}
\int d^3{\bf S}_{13}\int d^3{\bf S}_{24}\times\nonumber\\
& &\times\int{{d^3{\bf k}}\over{(2\pi)^3}}\int{{d^3{\bf p}}\over{(2\pi)^3}}
\int{{d^3{\bf q}}\over{(2\pi)^3}}e^{i{\bf k}\cdot({\bf S}_{24}-{\bf S}_2-{\bf S}_4)}
\times\nonumber\\
& &\times e^{i{\bf p}\cdot({\bf S}_{13}-{\bf S}_1-{\bf S}_3)}e^{i{\bf q}
\cdot({\bf S}-{\bf S}_{13}-{\bf S}_{24})}e^{-\beta H}\nonumber\\
& =&\Bigl({2\over{\pi}}\Bigr)^3\!\int_0^{\infty}\!S^2dS\int_0^{\infty}\!
S_{13}^2dS_{13}
\int_0^{\infty}\!S_{24}^2dS_{24}e^{-\beta H}\times\nonumber\\
& &\!\times\!\int_0^{\infty}\! dk{{\sin^2k\sin(kS_{24})}
\over{kS_{24}}}\!\int_0^{\infty}\!dp{{\sin^2p\sin(pS_{13})}\over{pS_{13}}}
\times\nonumber\\
& &\times\int_0^{\infty}dq{{\sin(qS)\sin(qS_{13})\sin(qS_{24})}\over{qSS_{13}S_{24}}}
\nonumber\\
& =&{1\over8}\!\int_0^2\!dS_{13}\int_0^2\!dS_{24}\int_{|S_{13}+S_{24}|}^{S_{13}+
S_{24}}\!SdS e^{\alpha(S^2-S_{13}^2-S_{24}^2)},\label{Z1}\\
& = &{1\over8}\overline{Z},\nonumber\\
\noalign{\rm where}\nonumber\\
\overline{Z}&=&\int_0^2dx{{\cosh(4\alpha x)-1}\over{2\alpha^2x}},\label{Z}
\end{eqnarray}
and $\alpha=\beta J/2$.  Eq. (\ref{Z}) was obtained
 previously, \cite{Luban} and an analysis of the integral
  was also presented.  The overall factor of ${1\over8}$ can be dropped, as it will not 
appear in the correlation functions.  In deriving $Z$, we note that the angles 
${\bf S}_{24}$ makes with ${\bf k}$ and ${\bf q}$ are independent, as are the angles 
${\bf S}_{13}$ makes with ${\bf k}$ and ${\bf p}$.  Hence, it is helpful to first 
integrate over the angles ${\bf S}_1$ and ${\bf S}_3$ make with ${\bf p}$ and the 
angles ${\bf S}_2$ and ${\bf S}_4$ make with ${\bf k}$.  Then, one may choose the 
coordinates of ${\bf k}$ and ${\bf p}$ relative to ${\bf S}_{24}$ and ${\bf S}_{13}$, 
respectively.

\subsection{Exact time evolution}

The dynamics of the spins arise from the Heisenberg equations of motion,
\begin{equation}
{{d{\bf S}_i}\over{dt}}={1\over{\tau}}\sum_{{j=1}\atop{<ij>}}^N{\bf S}_i
\times{\bf S}_j,\label{Sidot}
\end{equation}
where ${\bf S}_{N+i}={\bf S}_i$ for any integer $i$.  Our primary concern in
this paper is the case $N=4$, for which Eq. (\ref{Sidot}) may be rewritten as
\begin{eqnarray}
{{d{\bf S}_{1,3}}\over{dt}}& = &{1\over{\tau}}{\bf S}_{1,3}\times{\bf S}_{24},
\label{S1dot}\\
{{d{\bf S}_{2,4}}\over{dt}}& = &{1\over{\tau}}{\bf S}_{2,4}\times{\bf S}_{13}
\label{S2dot}\\
\noalign{\rm which lead to}\nonumber \\
{{d{\bf S}_{13}}\over{dt}}& = &{1\over{\tau}}{\bf S}_{13}\times{\bf S},\label{S13dot}\\
{{d{\bf S}_{24}}\over{dt}}& = &{1\over{\tau}}{\bf S}_{24}\times{\bf S},\label{S24dot}\\
\noalign{\rm and}\nonumber \\
{{d{\bf S}}\over{dt}}& = &0.\label{Sdot}
 \end{eqnarray}
The phenomenological classical spin precession rate ${1\over{\tau}}$ can be obtained from
first principles, starting from a quantum Heisenberg model whose classical counterpart is
given by the Hamiltonian in Eq. (1).  In that case, one simply obtains
$1/\tau = J/\hbar$. 
From Eq. (\ref{Sdot}) we see that the total spin ${\bf S}$ is conserved during the 
dynamical development, and is thus constant both in magnitude and direction. Equations 
(\ref{S13dot}) and (\ref{S24dot}) are easily interpreted as describing the precession 
of the vectors ${\bf S}_{13}$ and ${\bf S}_{24}$ about the constant vector ${\bf S}$, 
keeping their lengths invariant.  We note that Eqs. (\ref{S1dot}) and (\ref{S2dot}) 
show that each individual spin executes a more complicated dynamics, precessing about 
the particular ${\bf S}_{13}$ or ${\bf S}_{24}$ that describes the sum of its 
near-neighbor spins, which is itself precessing about the constant ${\bf S}$.  From  
well-known examples of rigid-body dynamics, we thus expect that the motion of the 
individual spin vectors will feature two frequencies, one for precession about
${\bf  S}$, 
and the other for precession about either ${\bf S}\pm {\bf S}_{24}$ or ${\bf
S}\pm {\bf S}_{13}$, respectively.
\begin{figure}
\vspace{0.3cm}
\epsfxsize=6.5cm
\centerline{\hspace{1cm}\epsffile{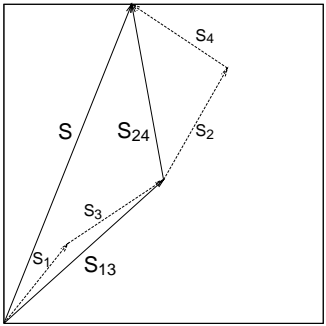}}
\vspace{1cm}
\caption{Sketch of the spin vectors.  ${\bf S}$, ${\bf S}_{13}$ and ${\bf S}_{24}$ 
all lie in a common plane where ${\bf S}_{13}$ and ${\bf S}_{24}$ each precess about ${\bf S}$.
   The unit vectors ${\bf S}_1$ and ${\bf S}_3$ initially point an arbitrary angle 
$\phi_{10}$ out of the plane, and precess about ${\bf S}_{13}$.  Similarly, the unit 
vectors ${\bf S}_2$ and ${\bf S}_4$   initially point an arbitrary angle $\phi_{20}$ 
out of the plane, and precess about ${\bf S}_{24}$.}
\label{fig2}  
\end{figure}

  The conservation of ${\bf S}$ enables us to solve Eqs. (\ref{S13dot}) and 
(\ref{S24dot}) exactly.  Since ${\bf S}={\bf S}_{13}+{\bf S}_{24}$ at all 
times, it is convenient to describe the motion in terms of the plane containing the 
three vectors ${\bf S}$, ${\bf S}_{13}$, and ${\bf S}_{24}$, as pictured in Fig. 2.  
We thus have
\begin{eqnarray}
{\bf S}_{24}(t)&=&C_{24}\hat{\bf s}+A_{24}[\hat{\bf x}\cos(st/\tau)-\hat{\bf y}
\sin(st/\tau)]\label{S24t}\\
\noalign{\rm and}\nonumber\\
{\bf S}_{13}(t)&=&C_{13}\hat{\bf s}+A_{13}[\hat{\bf x}\cos(st/\tau)-\hat{\bf y}
\sin(st/\tau)],
\end{eqnarray}
where $\hat{\bf s}$ is a unit vector parallel to ${\bf S}$, $\hat{\bf x}$ and 
$\hat{\bf y}$ are unit vectors normal to ${\bf S}$ satisfying $\hat{\bf x}\times
\hat{\bf y}=\hat{\bf s}$, thus completing the orthonormal basis set, and the constants 
$A_{ij}$ and $C_{ij}$  satisfy
\begin{eqnarray}
A_{13}^2+C_{13}^2&=&S_{13}^2\label{Pythagoras1}\\
\noalign{\rm and}\nonumber\\
A_{24}^2+C_{24}^2&=&S_{24}^2.\label{Pythagoras}
\end{eqnarray}
  We also have the relations
\begin{eqnarray}
A_{24}&=&-A_{13}\\
\noalign{\rm and}\nonumber\\
C_{13}+C_{24}&=S.\label{Csum}
\end{eqnarray}
By combining Eqs. (\ref{Pythagoras1}) and (\ref{Csum}), we obtain
\begin{eqnarray}
C_{13}&=&{{S^2+S_{13}^2-S_{24}^2}\over{2S}}\\
\noalign{\rm and}\nonumber\\
C_{24}&=&{{S^2+S_{24}^2-S_{13}^2}\over{2S}}.\label{Cvalues}
\end{eqnarray}

We now determine the individual spin vectors ${\bf S}_i$.  Because the four equations
in Eqs. (\ref{S1dot}) and (\ref{S2dot}) have the same general structure, it suffices 
to focus  
 on just one of them, say ${\bf S}_2(t)$.  
We write ${\bf S}_2$ in terms of its components, $S_{2s}$, $S_{2x}$, and $S_{2y}$, and 
make use of the standard Fourier transform $S_{2i}(t)=\int{{d\omega}\over{2\pi}}
\exp(i\omega t)S_{2i}(\omega)$.  We also let $S_{2\pm}=S_{2x}\pm iS_{2y}$, and 
$\omega_{\pm} =\omega\pm S/\tau$.
We then obtain
\begin{eqnarray}
\omega S_{2s}(\omega)&=&{{A_{13}}\over{2\tau}}\left[S_{2+}(\omega_{-})-S_{2-}
(\omega_{+})\right]\\
\noalign{\rm and}\nonumber\\
\omega S_{2\pm}(\omega)&=&\mp{{C_{13}}\over{\tau}}S_{2\pm}(\omega)\pm 
{{A_{13}}\over{\tau}}S_{2s}(\omega_{\pm}).
\label{S2FT}
\end{eqnarray}
Solving  for $S_{2\pm}(\omega)$, and then replacing $\omega$ by $\omega_{\mp}$,
 we have
\begin{equation}
S_{2\pm}(\omega_{\mp})={{A_{13}S_{2s}(\omega)}\over{C_{24}\pm\omega\tau}},
\label{S2pm}\end{equation}
where we have employed Eq. (\ref{Csum}).
Solving for $S_{2s}(\omega)$, we then find
\begin{equation}{{\omega}\over{C_{24}^2-(\omega\tau)^2}}\Big(S_{24}^2-(\omega\tau)^2
\Big)
S_{2s}(\omega)=0.
\end{equation}
This implies that $S_{2s}(\omega)$ has components from  three frequencies only, 
$\omega=0, \pm S_{24}/\tau$.  Thus, we write 
\begin{eqnarray}
S_{2s}(\omega)&=&2\pi S_{2s0}\delta(\omega)+\pi\Delta S_{2s0}
\Big[e^{i\phi_{20}}\delta(\omega-S_{24}/\tau)\nonumber\\
& &\hskip20pt +e^{-i\phi_{20}}\delta(\omega+S_{24}/\tau)\Big],\label{S2somega}\\
\noalign{\rm or in real time,}\nonumber\\
S_{2s}(t)&=&S_{2s0}+\Delta S_{2s0}\cos(S_{24}t/\tau+\phi_{20})\label{S2st},
\end{eqnarray}
where the constants $S_{2s0}$ and $\Delta S_{2s0}$ will be determined below, 
and $\phi_{20}$ is the arbitrary angle at which $S_2$  initially makes with the 
Fig. 2.  We note, however, that ${\bf S}_4$ must make the same initial angle 
$\phi_{40}=\phi_{20}$ with this plane, since both spins have unit length, and 
their sum  ${\bf S}_{24}$ is contained within that plane.  Analogously, 
${\bf S}_1$ and ${\bf S}_3$ both make the arbitrary initial angle $\phi_{10}$ with 
that plane.

Combining Eqs. (\ref{S2pm}) and (\ref{S2somega}), we may solve for 
$S_{2\pm}(\omega)$,
\begin{eqnarray}
S_{2\pm}(\omega)&=&{{2\pi A_{24}S_{2s0}\delta(\omega\pm S/\tau)}
\over{C_{24}}}\nonumber\\
&&+\pi A_{24}\Delta S_{2s0}\Bigl(e^{\pm i\phi_{20}}{{\delta[\omega
\pm(S-S_{24})/\tau]}\over{C_{24}-S_{24}}}\nonumber\\
& & +e^{\mp i\phi_{20}}{{\delta[\omega\pm(S+S_{24})/\tau]}
\over{C_{24}+S_{24}}}\Bigr).\label{S2pmomega}
\end{eqnarray}
Inverting the Fourier transform, and using $S_{2x}=(S_{2+}+S_{2-})/2$, 
$S_{2y}=(S_{2+}-S_{2-})/(2i)$, we have
\begin{eqnarray}
S_{2x}(t)&=&{{A_{24}}\over{C_{24}}}S_{2s0}\cos(St/\tau)\nonumber\\
&&+{{A_{24}\Delta S_{2s0}}\over{2[S_{24}+C_{24}]}}\cos[(S+S_{24})t/\tau+\phi_{20}]
\nonumber\\
& & -{{A_{24}\Delta S_{2s0}}\over{2[S_{24}-C_{24}]}}\cos[(S-S_{24})t/\tau-
\phi_{20}]\label{S2xt}\\
\noalign{\rm and}\nonumber\\
S_{2y}(t)&=&-{{A_{24}S_{2s0}}\over{C_{24}}}\sin(St/\tau)\nonumber\\
&&-{{A_{24}\Delta S_{2s0}}\over{2[S_{24}+C_{24}]}}\sin[(S+S_{24})t/\tau
+\phi_{20}] \nonumber\\ & &
+ {{A_{24}\Delta S_{2s0}}\over{2[S_{24}-C_{24}]}}\sin[(S-S_{24})t/\tau-
\phi_{20}].\label{S2yt}
\end{eqnarray}
Thus, as pictured in Fig. 2, ${\bf S}_{2}$, and correspondingly, ${\bf S}_4$,
precesses about ${\bf S}_{24}$, which
itself precesses about ${\bf S}$. As in Fig. 2,  ${\bf S}_1$ and ${\bf S}_3$
each precesses about ${\bf S}_{13}$, which itself precesses about ${\bf S}$.
 
  To evaluate the amplitudes $S_{2s0}$ amd $\Delta S_{2s0}$, we use the fact that 
$S_2^2(t)=1$ to   obtain 
\begin{equation}
S_{24}^2\Bigl({{S_{2s0}^2}\over{C_{24}^2}}+{{(\Delta S_{2s0})^2}\over{A_{24}^2}}
\Bigr)=1,\label{2unity}
\end{equation}
which is independent of $\phi_{20}$.  This equation provides a constraint upon the two amplitudes 
$S_{2s0}$ and $\Delta S_{2s0}$.  We may then immediately write down the analogous 
equation for the amplitudes $S_{4s0}$ and $\Delta S_{4s0}$ appearing in the analogous 
expression for ${\bf S}_4(t)$,
\begin{equation}
S_{24}^2\Bigl({{S_{4s0}^2}\over{C_{24}^2}}+{{(\Delta S_{4s0})^2}\over{A_{24}^2}}
\Bigr)=1.\label{4unity}
\end{equation}
Since ${\bf S}_4(t)={\bf S}_{24}(t)-{\bf S}_2(t)$, where ${\bf S}_{24}(t)$ is given 
in Eq. (\ref{S24t}), the expressions  for ${\bf S}_2(t)$ given by Eqs. (\ref{S2st}), 
(\ref{S2xt}), and (\ref{S2yt}), and the  analogous ones for  ${\bf S}_4(t)$ (with 
$S_{2s0}$ and $\Delta S_{2s0}$ replaced by $S_{4s0}$ and $\Delta S_{4s0}$, 
respectively) are consistent, provided that
\begin{eqnarray}
S_{4s0}&=&C_{24}-S_{2s0}\label{S2and4}\\
\noalign{\rm and}\nonumber\\
\Delta S_{4s0}&=&-\Delta S_{2s0}\label{DeltaS2and4}.
\end{eqnarray}
Substituting Eqs. (\ref{S2and4}) and ({\ref{DeltaS2and4}) into Eq. (\ref{4unity}),  
and subtracting the results from Eq. (\ref{2unity}), we have
\begin{equation}
S_{2s0}=S_{4s0}=C_{24}/2.\label{S2s0}
\end{equation}
Then, from Eq. (\ref{2unity}), we find
\begin{equation}
\Delta S_{2s0}=-\Delta S_{4s0}={{A_{24}}\over{S_{24}}}[1-S_{24}^2/4]^{1/2}.
\label{DeltaS2s0}
\end{equation} 
 In Eq. (\ref{DeltaS2s0}), we have made the arbitrary choice of assigning the positive 
sign to $\Delta S_{2s0}$, but that does not affect any of the results.  Thus, we have 
now completely determined the  dynamics of ${\bf S}_2(t)$ and ${\bf S}_4(t)$, except 
for the arbitrary phase $\phi_{20}$ representing the angle that ${\bf S}_2(0)$ makes 
with the plane containing ${\bf S}$, ${\bf S}_{24}$, and ${\bf S}_{13}$.  Similarly, 
${\bf S}_1(t)$ is obtained from Eqs. (\ref{S2st}), (\ref{S2xt}),  (\ref{S2yt}), 
(\ref{S2s0}), and (\ref{DeltaS2s0}) by replacing $A_{24}$, $C_{24}$,  $S_{24}$, and 
$\phi_{20}$ with $A_{13}$, $C_{13}$, $S_{13}$, and $\phi_{10}$, respectively.  
${\bf S}_3(t)$ is then obtained from ${\bf S}_1(t)$ in the same way as ${\bf S}_{4}(t)$ 
was obtained from ${\bf S}_2(t)$.

\section{Time Correlation Functions}

In this section, we utilize the exact results for the dynamics of the four spin vectors 
derived in the previous section to obtain analytical formulas for the three distinct 
time correlation functions.

\subsection{General Results}

 There are three inequivalent correlation functions, which we denote by ${\cal C}_{22}(t)=
\langle{\bf S}_2(t)\cdot{\bf S}_2(0)\rangle$, ${\cal C}_{12}(t)=\langle{\bf S}_1(t)
\cdot{\bf S}_2(0)\rangle$, and ${\cal C}_{24}(t)=\langle{\bf S}_2(t)\cdot{\bf S}_4(0)\rangle$, 
where $\langle\ldots\rangle={\rm Tr}[\exp(-\beta H)\ldots]/Z$.  These are the spin-spin 
autocorrelation function, the near-neighbor spin-spin correlation function, and the 
next-nearest-neighbor spin-spin correlation function, respectively.  In evaluating 
these functions, we must average over the initial conditions, which means not only the 
averages over $S$, $S_{24}$, and $S_{13}$, but also over the initial angles $\phi_{10}$ 
and $\phi_{20}$, which are present in Eq. (\ref{ZS}) in the integrations over the 
solid angles $\Omega_1$ and 
$\Omega_2$.   We note that all of the correlation functions depend upon the temperature 
through the parameter $\alpha$, but to keep the notation simple, we suppress that 
dependence.  The simplest of these ${\cal C}_{ij}(t)$ is ${\cal C}_{12}(t)$, for which the 
independent averages over $\phi_{10}$ and $\phi_{20}$ greatly simplify the final 
expression.  We find
\begin{eqnarray}
{\cal C}_{12}(t)&=&{1\over4}\langle C_{13}C_{24}+A_{13}A_{24}\cos(St/\tau)\rangle\\
& =&{1\over{4\overline{Z}}}\int_0^2dx\int_0^2dy\int_{|x-y|}^{x+y}s ds \> 
e^{\alpha(s^2-x^2-y^2)}\times\nonumber\\
& &\Bigl[{{s^4-(x^2-y^2)^2}\over{4s^2}}\nonumber\\
& &-\Bigl(y^2-{{[s^4+(x^2-y^2)^2]}\over{4s^2}}\Bigr)\cos(st/\tau)\Bigr],\label{C12t}
\end{eqnarray}
where $\alpha=\beta J/2$, $\overline{Z}$ is given by Eq. (\ref{Z}), and we have replaced $S$, 
$S_{13}$, and $S_{24}$ by $s$, $x$, and $y$, respectively.  After some algebra, the 
autocorrelation function is found to be
\begin{eqnarray}
{\cal C}_{22}(t)&=&I_0+I_1(t)+I_2(t)+I_3(t),\label{C22tinf}\\
\noalign{\rm where}\nonumber\\
I_0&=&{1\over4}\langle C_{24}^2\rangle,\label{tinfinite}\\
I_1(t)&=&{1\over4}\langle A_{24}^2\cos(St/\tau)\rangle,\label{I1}\\
I_2(t)&=&{1\over2}\langle{{A_{24}^2}\over{S_{24}^2}}[1-S_{24}^2/4]\cos(S_{24}t/\tau)
\rangle,\label{I2}\\
\noalign{\rm and}\nonumber\\
I_3(t)&=&{1\over2}\langle{{[1-S_{24}^2/4]}\over{S_{24}^2}}\Bigl([C_{24}^2+S_{24}^2]
\times\nonumber\\
& & \times\cos(St/\tau)\cos(S_{24}t/\tau)\nonumber\\
& & +2C_{24}S_{24}\sin(St/\tau)\sin(S_{24}t/\tau)\Bigr)\rangle.\label{I3}
\end{eqnarray}
Using the same notation as in Eq. (\ref{C12t}), we have
\begin{eqnarray}
I_0&=&{1\over{4\overline{Z}}}\int_0^2dx\int_0^2dy\int_{|x-y|}^{x+y}s ds\times
\nonumber\\
& &\hskip20pt\times e^{\alpha(s^2-x^2-y^2)}{{(s^2-x^2+y^2)^2}\over{4s^2}},\label{I0}\\
I_1(t)&=&{1\over{4\overline{Z}}}\int_0^2dx\int_0^2dy\int_{|x-y|}^{x+y}sds\> 
e^{\alpha(s^2-x^2-y^2)} \times\nonumber\\& & \hskip10pt\times\cos(st/\tau)
\Bigl(y^2-{{(s^2-x^2+y^2)^2}\over{4s^2}}\Bigr),\label{I1t}\\
I_2(t)&=&{1\over{2\overline{Z}}}\int_0^2dx\int_0^2dy{{(1-y^2/4)}\over{y^2}}
\cos(yt/\tau) \times\nonumber\\
& &\hskip30pt\times \int_{|x-y|}^{x+y}s ds \> e^{\alpha(s^2-x^2-y^2)}\times\nonumber\\ 
& &\hskip30pt\times \Bigl(y^2-{{(s^2-x^2+y^2)^2}\over{4s^2}}\Bigr),\\
\noalign{\rm and}\nonumber\\
I_3(t)&=&{1\over{2\overline{Z}}}\int_0^2dx\int_0^2dy{{(1-y^2/4)}\over{y^2}}
\int_{|x-y|}^{x+y}s ds \times\nonumber\\
& &\times e^{\alpha(s^2-x^2-y^2)}\Biggl[\Bigl(y^2+{{(s^2-x^2+y^2)^2}\over{4s^2}}
\Bigr)\times
\nonumber\\
 & &\hskip30pt\times \cos(yt/\tau)\cos(st/\tau)\nonumber\\
& &+{{y(s^2-x^2+y^2)}\over{s}}\sin(st/\tau)\sin(yt/\tau)\Biggr].\label{I3t}
\end{eqnarray}
Finally, ${\cal C}_{24}(t)$ may be found from
\begin{equation}
\langle {\bf S}_{24}(t)\cdot{\bf S}_{24}(0)\rangle=2{\cal C}_{22}(t)+2{\cal C}_{24}(t),
\end{equation}
or
\begin{eqnarray}
{\cal C}_{24}(t)&=&-{\cal C}_{22}(t)+{1\over2}\langle C_{24}^2+A_{24}^2\cos(St/\tau)\rangle
\nonumber\\
&=&2I_0+2I_1(t)-{\cal C}_{22}(t)\nonumber\\
&=&I_0+I_1(t)-I_2(t)-I_3(t).\label{C24tinf}
\end{eqnarray}

By interchanging the integrations over $x$ and $y$, it is easy to see that $\langle 
C_{13}C_{24}\rangle=\langle S^2/2-C_{24}^2\rangle$.
Thus, we note that ${\cal C}_{12}(t)$ is equivalent to
\begin{equation}
{\cal C}_{12}(t)
=\langle S^2\rangle/8-I_0-I_1(t).\label{C12tinf}
\end{equation}
We therefore remark that the ${\cal C}_{ij}(t)$ satisfy the conservation law,
\begin{equation}
{\cal C}_{22}(t)+{\cal C}_{24}(t)
+2{\cal C}_{12}(t)=\langle{\bf S}(t)\cdot{\bf S}(0)\rangle/4=\langle S^2\rangle/4,\label{s2}
\end{equation} 
a temperature-dependent quantity.
Hence, in the infinite time limit, two of the three correlation functions $C_{ij}(t)$ 
approach the same limit, 
\begin{equation}
{\rm lim}_{t\rightarrow\infty}{\cal C}_{22}(t)={\rm lim}_{t\rightarrow\infty}{\cal C}_{24}(t)=I_0,
\label{C22inf}
\end{equation}
but   
\begin{equation}
{\rm lim}_{t\rightarrow\infty}{\cal C}_{12}(t)=\langle S^2\rangle/8-I_0,\label{C12inf}
\end{equation}
since the other terms vanish due to the infinite number of oscillations of the 
integrand within the interval of integration. 
This is essentially a consequence of angular momentum conservation. \cite{deGennes}

\subsection{Reduction to Quadrature at Infinite Temperature}

\subsubsection{Analytic results at infinite temperature and time}

In the limits $t,T\rightarrow\infty$, we can evaluate the ${\cal C}_{ij}(t)$  analytically.  From Eqs. (\ref{C22tinf}),
(\ref{C24tinf}), and (\ref{C12tinf}), we note that these three functions are
all given by the $T\rightarrow\infty$ limit of $I_0$ and the  $I_i(t)$ for
$i=1,2,$ and 3. We first consider the simplest of these, $I_0$, which gives
the $t\rightarrow\infty$ limit.  
We find,
\begin{eqnarray}
\lim_{{t\rightarrow\infty}\atop{T\rightarrow\infty}}{\cal C}_{22}(t)=
\lim_{{t\rightarrow\infty}
\atop{T\rightarrow\infty}}{\cal C}_{24}(t)&=&{1\over4}+\delta_4
\label{C22infiniteT}\\
\noalign{\rm and}\nonumber\\
\lim_{{t\rightarrow\infty}\atop{T\rightarrow\infty}}{\cal C}_{12}(t)&=&
{1\over4}-\delta_4,
\label{C12infiniteT}
\end{eqnarray}
 where
\begin{equation}
\delta_4={8\over{45}}\ln 2-{{11}\over{180}}\approx 0.062115.\label{delta4}
\end{equation}

At first sight, one might have intuitively expected that the three ${\cal
C}_{ij}(t)$ should be equal to each other as $t,T\rightarrow\infty$, and
since $\lim_{T\rightarrow\infty}\langle S^2\rangle/4=1$,  Eq. (\ref{s2}) would
require each of them to  equal  $1/4$.  This  expectation is in fact
the result predicted by conventional diffusive spin dynamics in the infinite
temperature limit. \cite{LBC} Moreover, for finite times that formalism
predicts that all of the correlation functions depart from their common
infinite-time limit by terms that decay exponentially to zero.  However, our
present rigorous results, Eqs. (\ref{C22infiniteT}) - (\ref{delta4}),
as well as (\ref{C12longtime}) - (\ref{C24longtime}) in the following, show that these expectations are without
foundation.  Similar findings apply for the following simpler systems:  the
classical dimer, equilateral triangle, and regular tetrahedron, \cite{LBC},
for which the exact time correlation functions are derived as one-dimensional
integrals for all times and temperatures.

\subsubsection{One-dimensional integral representations}

In this subsection, we give one-dimensional integral representations for the
three time correlation functions at infinite temperature.  One important
advantage of these reduced forms is that they allow us to easily derive
analytical formulas for the leading corrections to the long-time asymptotic
values for finite times of each of the correlation functions.  Another
important advantage is that it becomes possible to obtain extremely accurate
numerical values for the infinite temperature correlation functions for all
times.  By comparison, for finite temperature, accurate numerical evaluation
of the three-dimensional integrals in Eqs. (\ref{I0}) - (\ref{I3t}) becomes a major
challenge.

We have found that  the three functions $I_i(t)$ may be written as

\begin{eqnarray}
\lim_{T\rightarrow\infty}I_1(t)&=&\int_0^2ds\>f_1(s)\cos(st^*)\nonumber\\
& &+\int_2^4ds\> g_1(s)\cos(st^*),\label{f1g1}\\
\lim_{T\rightarrow\infty} I_2(t)&=& \int_0^2ds f_2(s)\cos(st^*),\label{f2}\\
\noalign{\rm and}\nonumber\\
\lim_{T\rightarrow\infty} I_3(t)&=& \int_0^2ds f_3(s)\cos(st^*)\nonumber\\
& &+\int_2^4ds g_3(s)\cos(st^*)\nonumber\\
& &+\int_4^6ds h_3(s)\cos(st^*),\label{f3g3h3}    
\end{eqnarray}
where $t^{*}=t/\tau$, and analytic forms for the $f_i(s)$, $g_i(s)$ and $h_3(s)$ are listed in the
Appendix.  The infinite temperature correlation functions ${\cal C}_{22}(t)$,
${\cal C}_{24}(t)$, and ${\cal C}_{12}(t)$ are then simply found using
Eqs. (\ref{C22tinf}), (\ref{C24tinf}), and (\ref{C12tinf}), respectively.  Thus, we have reduced these infinite temperature correlation
functions to quadrature.  They are shown for $0\le t/\tau\le 10$ in Fig. 3a.

\begin{figure}
\vspace{0.3cm}
\epsfxsize=8cm
\centerline{\epsffile{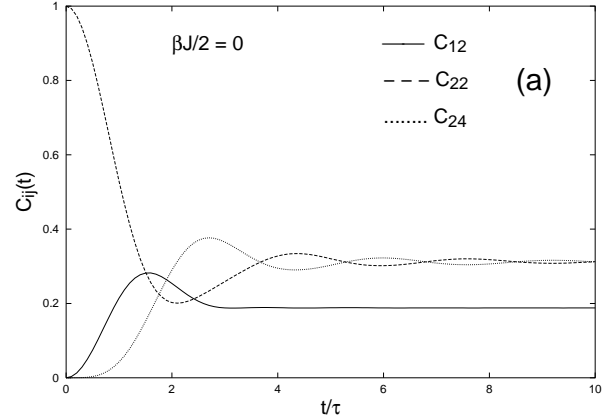}}
\vspace{0.3cm}
\centerline{\epsffile{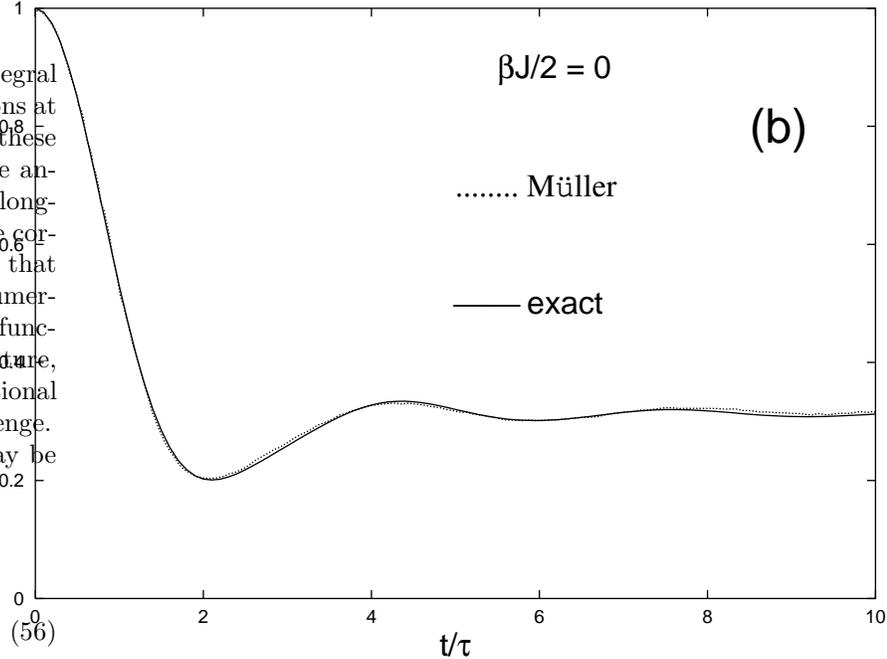}}
\vspace{0.3cm} 
\caption{(a) Plot of ${\cal C}_{12}(t)$ (solid), ${\cal C}_{24}(t)$ (dotted), and
${\cal C}_{22}(t)$ (dashed) versus $t/\tau$ in the infinite temperature
limit $\alpha=0$. (b) Comparison of the numerical results of M{\"u}ller [17] (dashed) with our
exact results (solid) for the infinite temperature ($\alpha=0$)
autocorrelation function ${\cal C}_{22}(t)$.}
\label{fig3}
\end{figure}

We remark that the infinite temperature autocorrelation function ${\cal C}_{22}(t)$ was obtained
previously using a purely numerical procedure. \cite{Mueller} In Fig. 3b, we
have compared those published results with our exact formula at infinite
temperature.  Although there was some distortion in the axes in the published
figure, using a pure rotation to account for this distortion led to the
excellent agreement between the numerical and exact results.
 
From Fig. 3a, the autocorrelation function ${\cal C}_{22}(t)$ decreases from its initial 
value ${\cal C}_{22}(0) = 1$,
 then undershoots its aymptotic limit ${1\over{4}}+\delta_4$, and approaches
this limit by oscillating
about it for a rather long time.  On the other hand, spins of different sites
are initially uncorrelated at infinite temperature,
 ${\cal C}_{12}(0)={\cal C}_{24}(0)=0$. At later limes, these functions
both overshoot their respective asymptotic limits ${1\over{4}}-\delta_4$ and
${1\over{4}}+\delta_4$,  and then oscillate about them.
The oscillations of ${\cal C}_{12}(t)$ decay so rapidly that they are
barely discernible in this figure.  On the other hand the oscillations of
${\cal C}_{24}(t)$ are of the same amplitude and persist as long as do those of
${\cal C}_{22}(t)$, and are likewise easily seen in this figure.  In
addition, after ${\cal C}_{22}(t)$ and ${\cal C}_{24}(t)$ first become equal
to each other, they braid about each other in their approaches to the same
aymptotic limit.

In order to see more clearly how this occurs, we have found analytic 
expressions for
the leading behaviors of the correlation functions for long times, $t>>\tau$.
We first consider ${\cal C}_{12}(t)$. In this case, besides the constant
$I_0$, we only need to evaluate $I_1(t)$.  To do so,  we integrate both terms
in Eq. (\ref{f1g1}) by parts, treating  $\cos(st^{*})$ as the variable to be
integrated, and $f_1(s)$ and $g_1(s)$ as the variables to be differentiated.
From the results in the Appendix, it is seen that $f_1(s)$ and $g_1(s)$ as
well as their first three derivatives are continuous at $s=2$.  In
addition, since the relevant integration endpoint values and derivatives at 
$s=0$ and $s=4$
also make no contribution through third order in the repeated integrations by 
parts,
the leading contribution to the aymptotic behavior arise from the
non-vanishing $f_1^{'''}(0)$ and $g_1^{'''}(4)$. The final result for the
leading behavior is given by 
\begin{equation}
\lim_{{T\rightarrow\infty}\atop{t>>\tau}}{\cal C}_{12}(t)\rightarrow
 {1\over4}-\delta_4
+{1\over{4t^{*4}}}[{{3}\over{4}}-\cos(4t^{*})].\label{C12longtime}
\end{equation}

On the other hand, ${\cal C}_{22}(t)$ and ${\cal C}_{24}(t)$ at infinite
temperature also depend upon $I_2(t)$ and $I_3(t)$.  Again, we integrate by
parts in a similar fashion, treating 
$f_2(s)$, $f_3(s)$, $g_3(s)$, and $h_3(s)$ as the variables to be differentiated.  The leading non-vanishing
contributions to ${\cal C}_{22}(t)$ and ${\cal C}_{24}(t)$ from these integrations by parts are both of
second order.  For $I_2(t)$, the leading non-vanishing contribution comes from the
non-vanishing $f_2^{'}(0)$ and $f_2^{'}(2)$, the latter of which is a
non-trivial number.  For $I_3(t)$, the leading non-vanishing contribution  
arises from $f_3^{'}(0)$, $f_3^{'}(2)$ and
$g_3^{'}(2)$. Although both $f_3(s)$ and $g_3(s)$ have non-trivial values and
derivatives at their matching point $s=2$,  the difference
between  their derivatives is a trivial, but non-vanishing value. In addition,
the functions $f_3(s)$ and $g_3(s)$  both have non-trivial
values and derivatives at their matching point $s=4$, but these values and
derivatives are 
equal, and thus their contribution in second order to the integration by parts
vanishes.
We thus obtain the long-time behaviors at infinite temperature,
 
\begin{eqnarray}
\lim_{{T\rightarrow\infty}\atop{t>>\tau}}{\cal C}_{22}(t)&\rightarrow&
{1\over4}+\delta_4\nonumber\\
& &+{{1}\over{60t^{*2}}}[5-(29+8\ln 2)\cos(2t^{*})],
\label{C22longtime}\\
\noalign{\rm and}\nonumber\\
\lim_{{T\rightarrow\infty}\atop{t>>\tau}}{\cal C}_{24}(t)&\rightarrow&
{1\over4}+\delta_4\nonumber\\
& &-{{1}\over{60t^{*2}}}[5-(29+8\ln 2)\cos(2t^{*})].\label{C24longtime}
\end{eqnarray}

We  note that ${\cal C}_{12}(t)$ decays much more rapidly 
($\propto 1/t^{*4}$) to its constant
long-time limit than do either ${\cal C}_{22}(t)$ or ${\cal C}_{24}(t)$
 ($\propto 1/t^{*2}$).  The long-time  
braiding of these functions about each other  arises from
 the opposite signs of their oscillatory terms. Furthermore, at long times, 
${\cal C}_{12}(t)$ oscillates
with twice the frequency of the long-time oscillations of ${\cal C}_{22}(t)$ and 
${\cal C}_{24}(t)$.

\subsection{Results for finite temperatures}

At finite $T$, we evaluate $I_0(\alpha)$ and $\langle S^2\rangle  
(\alpha)$ numerically.  For 
$|\alpha|<1$,  breaking each integral into 100 intervals is sufficient to obtain 
0.1\% accuracy. Note that this means there are $10^6$ integration intervals overall.  
However, for low $T$ ($|\alpha|>1$), the number of intervals necessary to obtain that 
degree of accuracy increases.  At $|\alpha|=10$, one needs to break up each
integration domain 
into 400 intervals, for instance.  In Fig. 4, we have plotted the infinite-time limit 
of the spin-spin correlation functions $I_0(\alpha)$ and $\langle S^2\rangle/8-I_0(\alpha)$ for both the 
ferromagnetic (FM) and antiferromagnetic (AFM) cases.  As $\alpha\rightarrow0$, one 
obtains the analytic limits given by Eqs. (\ref{C22infiniteT}) and (\ref{C12infiniteT}).  
However, in the low $T$ limit $|\alpha|\rightarrow\infty$, both $I_0$ and $\langle S^2\rangle/8-I_0\rightarrow 0 (1)$ 
for 
the AFM (FM) case, respectively.  This just tells us that at $T=0$, all of the spins 
are 
aligned in the FM case, and in the AFM case, their sum is 0.  We note that for 
$|\alpha|<<1$, $I_0(\alpha)$ obeys the inversion symmetry, equivalent to 
$\partial{I_0}/\partial\alpha\Big|_{\alpha=0}$ exists.  We note that for the
AFM case. $\lim_{t\rightarrow\infty}{\cal C}_{12}(t)$ is negative for 
$\alpha<-0.71$.  This just reflects the fact that for the antiferromagnetic
ring, the spins on neighboring sites are anticorrelated at long times and for
temperatures that are not too large.

\begin{figure}
\vspace{0.3cm}
\epsfxsize=8cm
\centerline{\epsffile{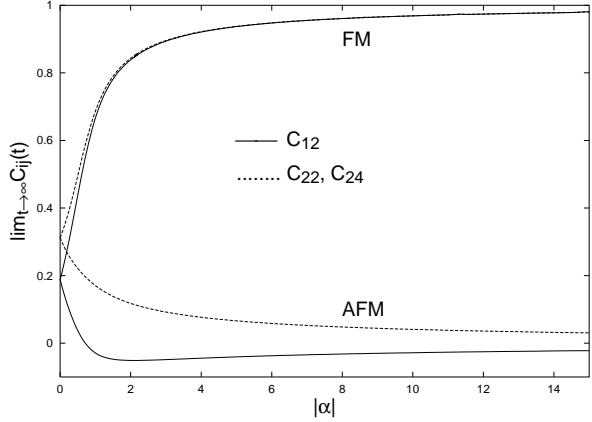}}
\vskip0.3cm
\caption{Plot of $I_0(\alpha)={\rm lim}_{t\rightarrow\infty}C_{22}(t)=
{\rm lim}_{t\rightarrow\infty}C_{24}(t)$ and $\langle S^2\rangle/8-I_0(\alpha)=
{\rm lim}_{t\rightarrow\infty}C_{12}(t)$, as a function of
 $|\alpha|$, for the FM ($\alpha>0$) and AFM ($\alpha<0$) cases.}\label{fig4}
\end{figure}

 At finite $T$  we also may evaluate  the $I_i(t)$ numerically from the triple
 integral forms, Eqs. (\ref{I1t}) - (\ref{I3t}).
As for $I_0$, we break each of the three integrals into $N$ intervals.  At the lowest
 $T$ values considered ($\alpha=-20$), it is necessary to take $N\ge1000$ to
 achieve sufficient accuracy.  In Figs. 5-7, we have plotted the ${\cal C}_{ij}(t)$ for 
the FM case with $\alpha=0.5,  2, $ and 10, and compared with the analytic
 results for $\alpha=0$.
 In each of these figures, ${\cal C}_{12}(t)$ decays to the 
equilibrium 
value $\langle S^2\rangle/8-I_0(\alpha)$  more rapidly than  ${\cal C}_{22}(t)$ and ${\cal C}_{24}(t)$ decay to 
their mutual 
equilibrium value $I_0(\alpha)$, while
 oscillating for a few periods about the latter.   As $T$ decreases, all of the ${\cal
 C}_{ij}(0)$ increase monotonically, approaching unity as $T\rightarrow0$.  In
 addition, the oscillations persist to much longer times.   Also, as seen in
 Figs. 5 and 8, as  $T$ decreases,
${\cal C}_{12}(t)$ 
 oscillates for an increasing amount of time, and ${\cal C}_{22}(t)$ and ${\cal C}_{24}(t)$ 
oscillate about it for an even longer period of time. 

\begin{figure}
\vspace{0.3cm}
\epsfxsize=8cm
\centerline{\epsffile{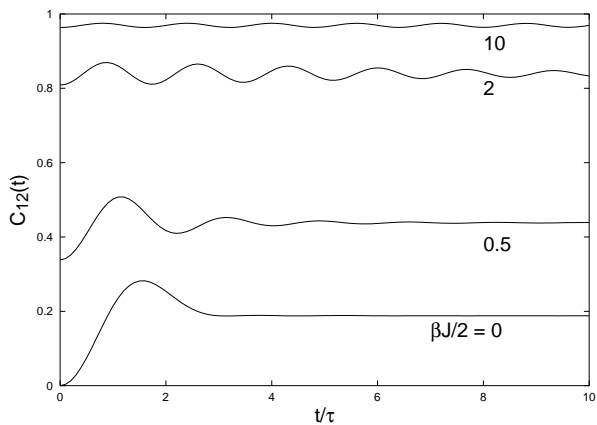}}
\vskip0.3cm
\caption{Plots of  ${\cal C}_{12}(t)$  
versus $t/\tau$ for the   $\alpha=\beta J/2$ FM cases 0, 0.5, 2, and 10, as indicated.}\label{fig5}
\end{figure}

\begin{figure}
\vspace{0.3cm}
\epsfxsize=8cm
\centerline{\epsffile{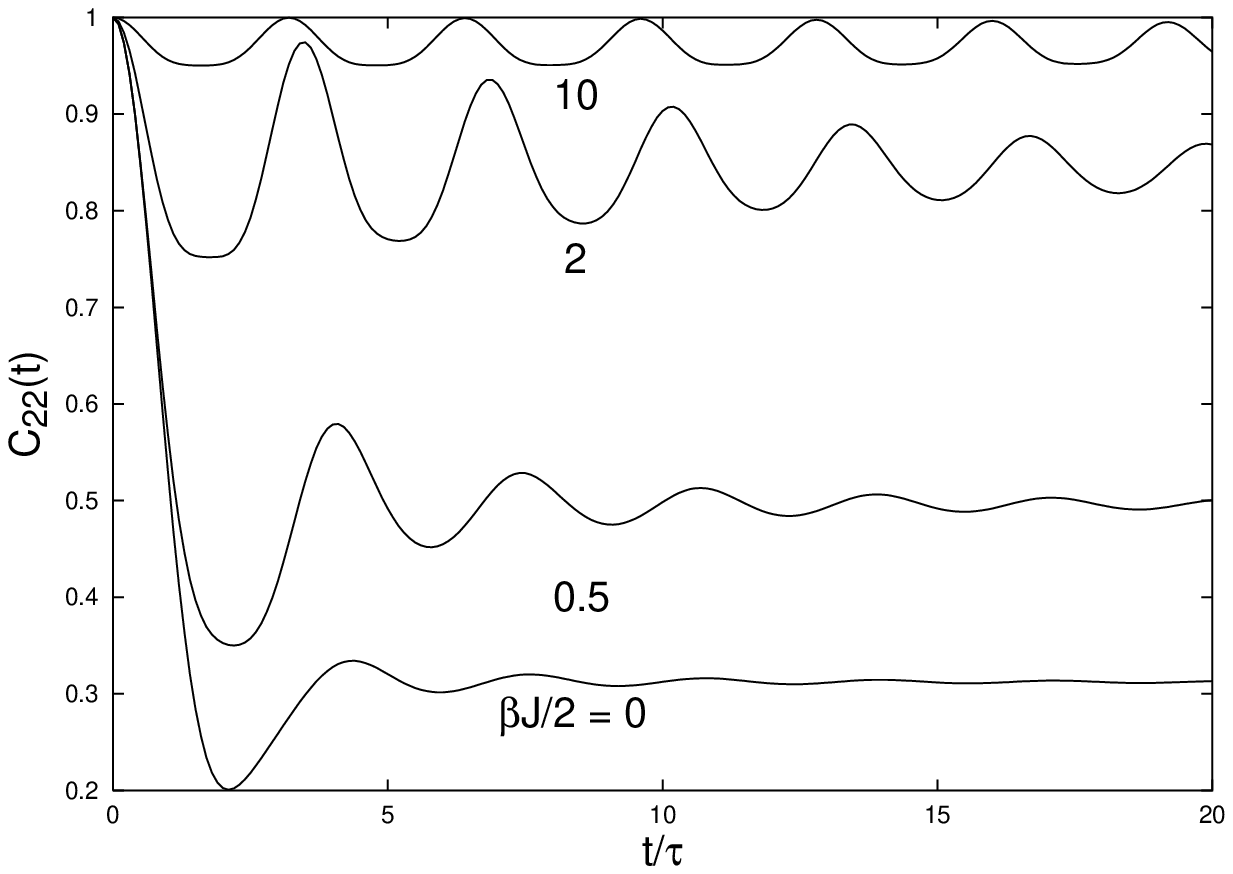}}
\vskip0.3cm
\caption{Plots of  ${\cal C}_{22}(t)$  
versus $t/\tau$ for the $\alpha=\beta J/2$
FM cases 0, 0.5, 2, and 10, as indicated.}\label{fig6}
\end{figure}

\begin{figure}
\vspace{0.3cm}
\epsfxsize=8cm
\centerline{\epsffile{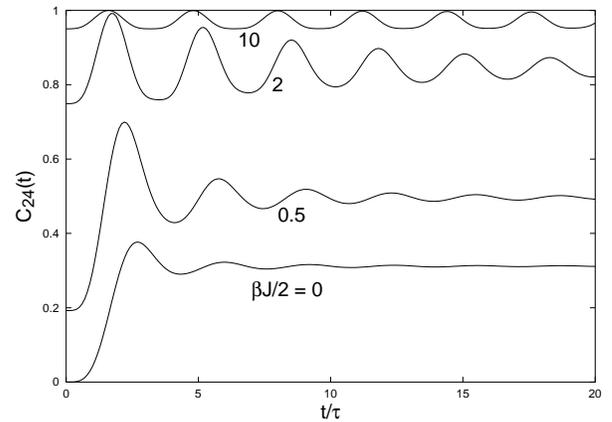}}
\vskip0.3cm
\caption{Plots of  ${\cal C}_{24}(t)$ 
versus $t/\tau$ for the  $\alpha=\beta J/2$ FM cases 0, 0.5, 2, and 10, as indicated.}\label{fig7}
\end{figure}

 At lower
 $T$, the amplitudes of the oscillations eventually reach a maximum, so that
 the oscillations in ${\cal C}_{12}(t)$ for $\alpha=0.5,2,10$ are distinctly
 noticeable.  As $T\rightarrow0$, the lifetimes of the oscillations appear to
 diverge,
 but their amplitudes become vanishingly small. At $\alpha=10$, we have shown
 the behaviors of ${\cal C}_{12}(t)$ (solid), ${\cal C}_{22}(t)$ (dashed), and
 ${\cal C}_{24}(t)$ (dotted) together in Fig. 8 for the extended time
 domain $0\le t/\tau\le 30$.   
Throughout this domain, the decay of the oscillations in all three 
 correlation 
functions is small but discernible.  However, careful inspection of the  oscillating waveforms 
reveals that ${\cal C}_{12}(t)$  oscillates with twice the frequency of the
 other two, continuing the pattern that we have already seen for infinite 
temperature.
   Note that ${\cal C}_{12}(t)$ 
appears to oscillate nearly as a simple cosine function, but ${\cal C}_{22}(t)$ and 
${\cal C}_{24}(t)$ have  a more complicated oscillatory behavior, with a fundamental frequency 
that is 
one-half that of ${\cal C}_{12}(t)$, and they are almost completely out of phase with respect 
to one another.

\begin{figure}
\vspace{0.3cm}
\epsfxsize=8cm
\centerline{\epsffile{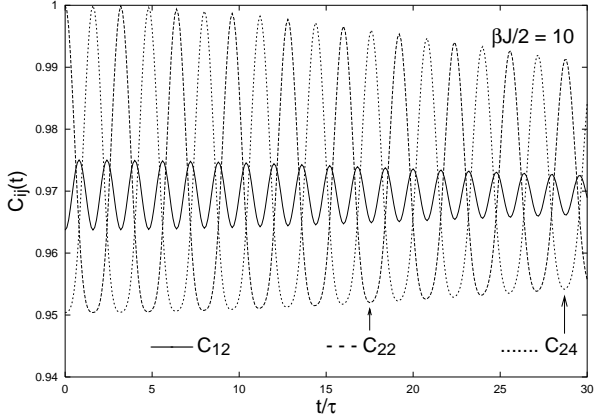}}
\vskip0.3cm
\caption{Plot of ${\cal C}_{12}(t)$ (solid), ${\cal C}_{24}(t)$ (dotted), and
${\cal C}_{22}(t)$ (dashed)
 versus $t/\tau$ for the low temperature FM case $\alpha=\beta J/2=10$. Note
that ${\cal C}_{22}(0)=1$ and ${\cal C}_{24}(0)\approx0.95$.}\label{fig8}
\end{figure}

The corresponding results  for the AMF case are shown in Figs. 9-11, 
for which $\alpha=$ -0.5,  -2, and - 20, respectively.  The last of
 these, $\alpha=-20$, took weeks of computational time to obtain sufficient accuracy.  
In these cases, we presented the ${\cal C}_{12}(t)$, ${\cal C}_{22}(t)$, and ${\cal C}_{24}(t)$ 
data as solid, dashed, and dotted curves, respectively.  We note that as the temperature is 
lowered, ${\cal C}_{12}(0)$ decreases towards the value -1, which would correspond to perfect
 AFM behavior.  However, ${\cal C}_{12}(t)$ then increases with $t$, reaches a maximum, and 
then decreases to the asymptotic , infinite time limit.  In addition, as $T$ is lowered,
 ${\cal C}_{24}(0)$ increases towards +1, approaching ${\cal C}_{22}(0)$.  Then, at some time $t_1$, 
${\cal C}_{22}(t_1)$ first equals ${\cal C}_{24}(t_1)$, and thereafter, the two functions are braided 
about each other.  The braiding oscillations decrease in amplitude as $T$ is decreased,
 so that the overall ${\cal C}_{ij}(t)$ all approach non-oscillatory uniform curves as 
$T\rightarrow0$. 

\begin{figure}
\vspace{0.3cm}
\epsfxsize=8cm
\centerline{\epsffile{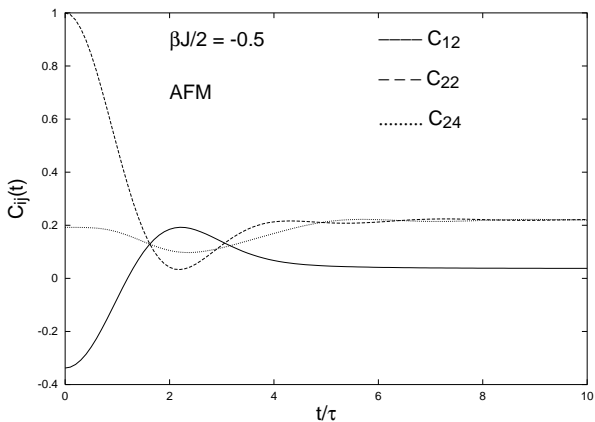}}
\vskip0.3cm
\caption{ Plot of ${\cal C}_{12}(t)$ (solid), ${\cal C}_{24}(t)$ (dotted), and
${\cal C}_{22}(t)$ (dashed) 
versus $t/\tau$ for the AFM case $\alpha=-0.5$.}\label{fig9}
\end{figure}

\begin{figure}
\vspace{0.3cm}
\epsfxsize=8cm
\centerline{\epsffile{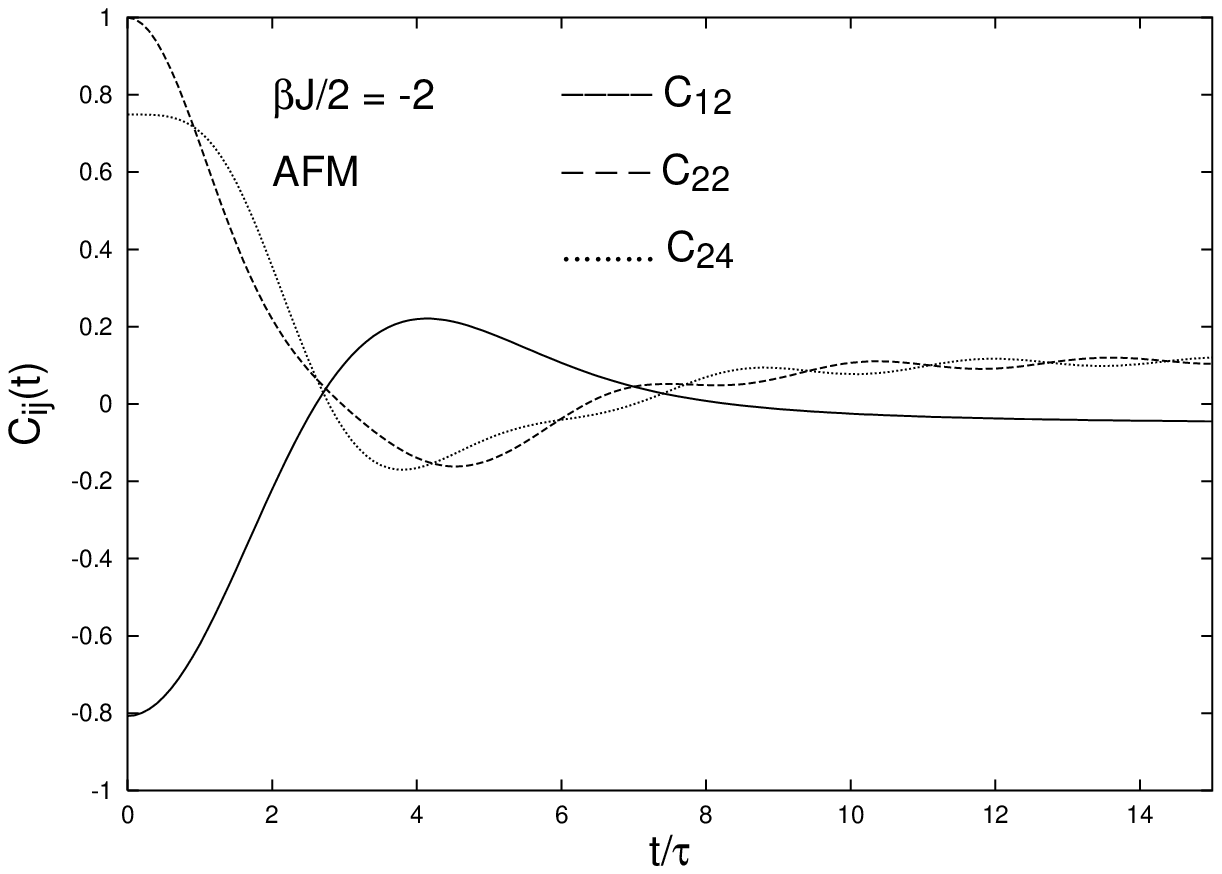}}
\vskip0.3cm
\caption{Plot of ${\cal C}_{12}(t)$ (solid), ${\cal C}_{24}(t)$ (dotted), and
${\cal C}_{22}(t)$ (dashed) 
versus $t/\tau$ for the AFM case $\alpha=-2$.}\label{fig10}
\end{figure}

\begin{figure}
\vspace{0.3cm}
\epsfxsize=8cm
\centerline{\epsffile{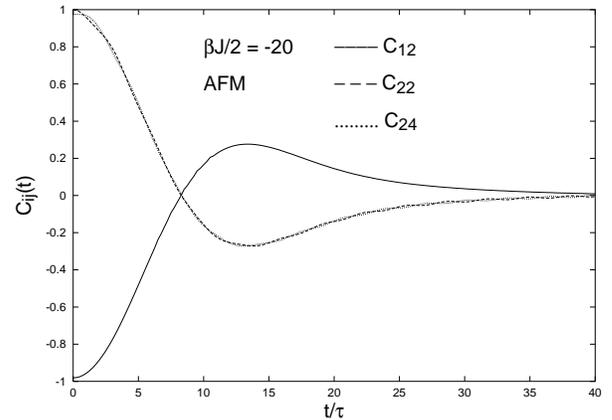}}
\vskip0.3cm
\caption{Plot of  ${\cal C}_{12}(t)$ (solid), ${\cal C}_{24}(t)$ (dotted), and
${\cal C}_{22}(t)$
(dashed) versus $t/\tau$ for  the very low temperature AFM case $\alpha=-20$.}
\label{fig11}
\end{figure}

From numerical simulation studies of more complicated mesoscopic classical
systems, it has been 
suggested that the low temperature AFM autocorrelation function should scale, 
approaching uniform functions of $tT^{1/2}$. \cite{Schroeder}  To investigate whether 
such a scenario 
holds for this exactly solved four-spin system, we first plotted the AFM autocorrelation
 function at the low $T$ values we considered.  This is shown in Fig. 12.   Although 
there are oscillations that persist to increasing times as $T$ is lowered, the overall 
shape of the curves does not change its shape qualitatively, suggesting that $C_{22}(t)$
 might 
indeed scale as a single function of $tT^{1/2}$ as $T\rightarrow0$.  In fact,
this behavior has been established by analytical means for the simpler cases
of the classical dimer, equilateral triangle, and regular tetrahedron. \cite{Luban,LBC} 

\begin{figure}
\vspace{0.3cm}
\epsfxsize=8cm
\centerline{\epsffile{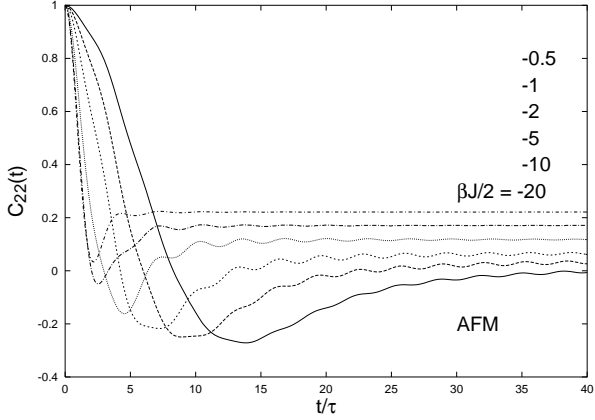}}
\vskip0.3cm
\caption{Plot of  ${\cal C}_{22}(t)$  for the $\alpha=\beta J/2$ AFM cases
-0.5, -1, -2, -5, -10, and -20, correspondingly from top to bottom at large $t/\tau$, as indicated.}\label{fig12}
\end{figure}

In Fig. 13, we therefore plotted ${\cal C}_{22}(t)$ versus 
$t/[\tau|\beta J/2|^{1/2}]$, which is proportional to
 $tT^{1/2}$, to check this notion quantitatively.  Indeed, 
the curves do scale, except for the braiding oscillations, which
 are decreasing in magnitude as $T$ decreases.  Thus, curves for the lowest two temperatures, 
$\alpha=-10$ and $\alpha=-20$, would nearly fall on top of each other
 if the oscillations were not present.  Similar low temperature scaling behavior of 
${\cal C}_{24}(t)$ is shown in Fig. 14, which also includes the braiding oscillations.  
$C_{12}(t)$ exhibits a clearer example of the scaling, as shown in Fig. 15, since it 
does not contain any braiding oscillations.    The major deviation from scaling occurs 
at very short times, although the differences between the curves at $\alpha=-10$ and 
$\alpha=-20$ are not so large there.  Since  ${\cal
C}_{12}(0) = -1$ in the zero temperature limit, this deviation from scaling probably arises from the fact that 
the $T=0$ limit has not yet been reached for such short times.

\begin{figure}
\vspace{0.3cm}
\epsfxsize=8cm
\centerline{\epsffile{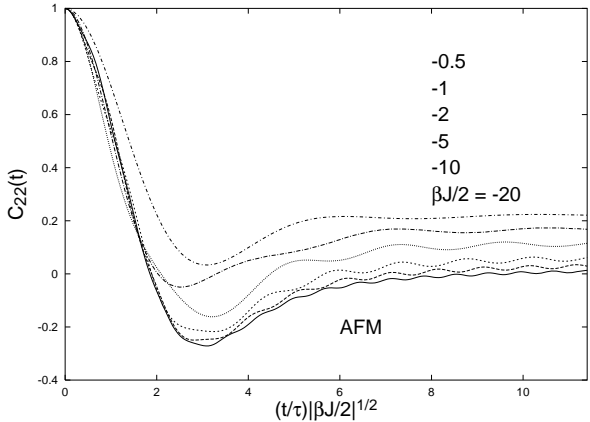}}
\vskip0.3cm
\caption{Plot of  ${\cal C}_{22}(t)$  as a function of the scaled time
$(t/\tau)|\beta J/2|^{1/2}$, for the $\alpha=\beta J/2$ AFM cases
-0.5, -1, -2, -5, -10, and -20, correspondingly from top to bottom at large $t/\tau$, as indicated.}\label{fig13}
\end{figure}

\begin{figure}
\vspace{0.3cm}
\epsfxsize=8cm
\centerline{\epsffile{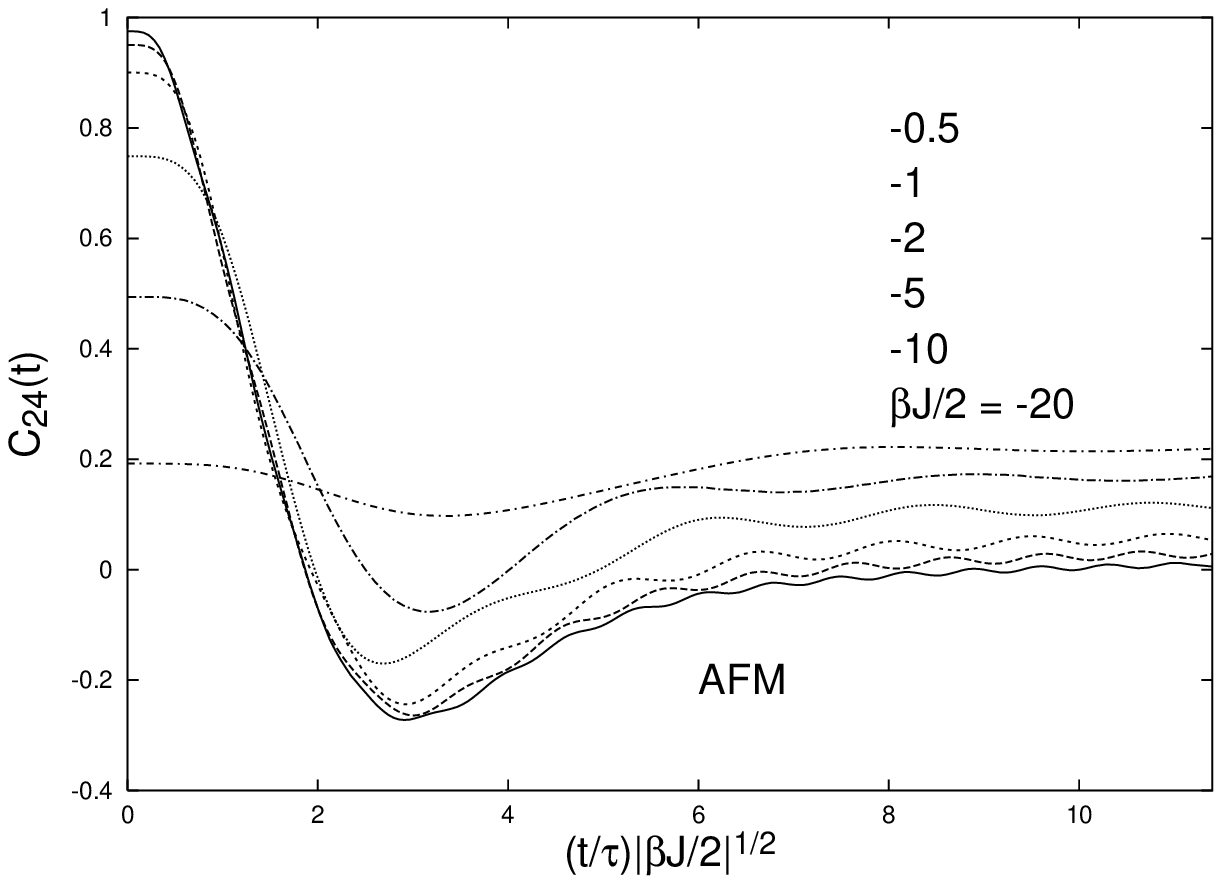}}
\vskip0.3cm
\caption{Plot of  ${\cal C}_{24}(t)$  as a function of the scaled time
$(t/\tau)|\beta J/2|^{1/2}$, for the $\alpha=\beta J/2$ AFM cases
-0.5, -1, -2, -5, -10, and -20, correspondingly from top to bottom at large $t/\tau$, as indicated.}\label{fig14}
\end{figure}

\begin{figure}
\vspace{0.3cm}
\epsfxsize=8cm
\centerline{\epsffile{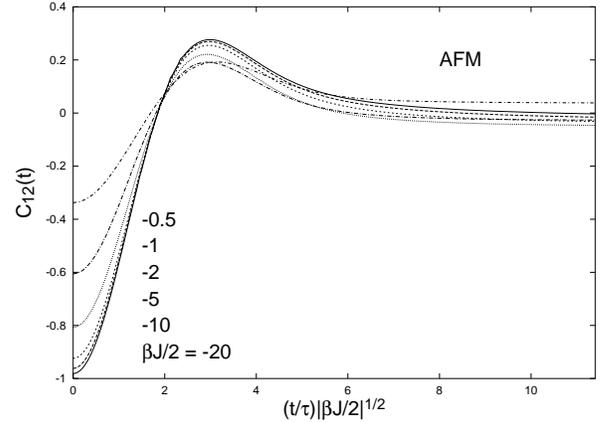}}
\vskip0.3cm
\caption{Plot of  ${\cal C}_{12}(t)$  as a function of the scaled time
$(t/\tau)|\beta J/2|^{1/2}$, for the $\alpha=\beta J/2$ AFM cases
-0.5, -1, -2, -5, -10, and -20, correspondingly from top to bottom at small $t/\tau$, as indicated.}\label{fig15}
\end{figure}

At higher temperatures, the scaling property gradually breaks down.  This is clearly seen for
$\alpha=-0.5$ in Figs. 14 at 15, for which the short-time values of ${\cal
C}_{24}(t)$ and ${\cal C}_{12}(t)$ deviate greatly from the values (1 and -1)
obtained respectively in the low temperature limit $\alpha\rightarrow
-\infty$.  For ${\cal C}_{12}(t)$, the deviations are also rather large at long
times.  However, for all three correlation functions, even at
$\alpha=-0.5$, the positions of the dip for ${\cal C}_{22}(t)$ and ${\cal C}_{24}(t)$ and the
peak for ${\cal C}_{12}(t)$ still scale. 

\subsection{Fourier Transforms}

We now evaluate the Fourier transforms $\tilde{I}_i(\omega)$ of the $I_i(t)$.  
Since  ${\cal C}_{22}(t)$ and ${\cal C}_{24}(t)$ both approch the constant $I_0\ne0$ as 
$t\rightarrow\infty$, and 
 ${\cal C}_{12}(t)$ approaches $\langle S^2\rangle/8-I_0\ne0$ in the same limit,
 the $I_0\ne0$ or $\langle S^2\rangle/8-I_0\ne0$ 
present in the respective ${\cal C}_{ij}(t)$ give rise to  delta functions in terms of the 
angular 
frequency $\omega$, in $\tilde{{\cal C}}_{ij}(\omega)$, equal to either 
$2\pi I_0\delta(\omega)$ or $2\pi(\langle s^2\rangle/8-I_0)\delta(\omega)$, which can be written down 
by inspection from Eq. (\ref{I0}).  We shall therefore evaluate the Fourier 
transform of the deviations $\delta {\cal C}_{ij}(t)=
{\cal C}_{ij}(t)-{\rm lim}_{t\rightarrow\infty}{\cal C}_{ij}(t)$,
\begin{eqnarray}
\tilde{{\cal C}}_{ij}(\omega)&=&\int_{-\infty}^{\infty}dt e^{-i\omega t}{\cal C}_{ij}(t)
\nonumber\\
&=&2\pi I_0\delta(\omega)+\delta\tilde{{\cal C}}_{ij}(\omega)\\
\noalign{\rm for $(i,j)$=(2,2) and (2,4), and}\nonumber\\
\tilde{{\cal C}}_{12}(\omega)&=&2\pi(\langle S^2\rangle/8-I_0)\delta(\omega)+\delta\tilde{{\cal C}}_{12}(\omega),\\
\noalign{\rm  where}\nonumber\\
\delta\tilde{{\cal C}}_{ij}(\omega)&=&\int_{-\infty}^{\infty}dt e^{-i\omega t}
\delta{\cal C}_{ij}(t)  .\label{Ftdef}
\end{eqnarray}

Causality requires that  $\delta\tilde{{\cal C}}_{ij}(-\omega)=
\delta\tilde{{\cal C}}_{ij}(\omega)$.  Thus, it suffices to evaluate the 
$\delta\tilde{{\cal C}}_{ij}(\omega)$ for $\omega\ge0$. For simplicity, we first evaluate 
$\tilde{I}_2(\omega)$, and let $\tilde{\omega}=\omega\tau$.  Since the only 
time-dependence within the expression for $I_2(t)$ appears in the factor 
$\cos(yt/\tau)$, Fourier transformation replaces this factor with 
$[\delta(\omega-y/\tau)+\delta(\omega+y/\tau)]/2$, which can be written as 
$(\tau/2)[\delta(y-\tilde{\omega})+\delta(y+\tilde{\omega})]$.  The second 
$\delta$-function does not contribute to the Fourier transform for $\omega>0$, 
so we obtain for 
$\omega\ge0$,
\begin{eqnarray}
\tilde{I}_2(\omega)&=& {{\pi\tau\Theta(2-\tilde{\omega})(1-\tilde{\omega}^2/4)}
\over{8\tilde{\omega}\overline{Z}}}\int_0^2xdx
\int_{-1}^1dz e^{2\alpha\tilde{\omega}xz}\times\nonumber\\
& &\times \Bigl(x^2+\tilde{\omega}^2-2xz\tilde{\omega}-{{(x^2-\tilde{\omega}^2)^2}
\over{x^2+\tilde{\omega}^2+2xz\tilde{\omega}}}\Bigr),\label{I2omega}
\end{eqnarray}
where $\Theta(x)$ is the Heaviside step function, and we have made the change of 
variables $s^2\rightarrow  {x}^2+\tilde{\omega}^2+2xz\tilde{\omega}$ for ease of 
computation.

We now evaluate $\tilde{I}_1(\omega)$.  As in the expression for 
$\tilde{I}_2(\omega)$, Fourier transformation of the factor $\cos(st/\tau)$ replaces 
it with $(\tau/2)[\delta(s-\tilde{\omega})+\delta(s+\tilde{\omega})]$, and the second 
term does not contribute to the integrals for $\omega>0$.  However, since $0\le s\le 4$, 
there are now two regions of integration over the variables $x$ and $y$. For 
$2\le\tilde{\omega}\le4$, the only region of integration is the interior of
the 
 isosceles triangle 
with sides obeying $x=2$, $y=2$, and 
$x+y=\tilde{\omega}$, and corners at their intersections.  For $0\le\tilde{\omega}\le2$, the region of integration is 
the interior of the pentagon with sides obeying $y=2$, $x=2$, $y-x=-\tilde{\omega}$, 
$y+x=\tilde{\omega}$, and $y=x+\tilde{\omega}$. This interior region is
symmetric 
about the line 
$y=x$, and can be broken up into two regions of integration.  
The first region is the interior of a rectangle
 rotated $45^{\circ}$ about the axis normal to the $xy$ plane, with sides obeying 
$y=x\pm\tilde{\omega}$ and $x+y=2\pm(2-\tilde{\omega})$.  The second region is the 
interior of the isosceles triangle with sides obeying  $y=2$, $x=2$, and 
$y=-x+4-\tilde{\omega}$. 

 In the triangular integration regions, we maintain the integration variables $x$ and 
$y$, keeping account of the integration limits.  However, in the rectangular 
integration region, it is convenient to perform a rotation of the axes by 
$45^{\circ}$, letting $r=x-y$, $s=x+y$, and incorporating the Jacobian, which 
replaces the differential integration area $dx dy$ with $dr ds/2$.  We thus have
\begin{eqnarray}
\tilde{I}_1(\omega)&=&{{\pi\tau\Theta(\tilde{\omega}-2)\Theta(4-\tilde{\omega})}
\over{16\tilde{\omega}\overline{Z}}}\int_{\tilde{\omega}-2}^2dx
\int_{\tilde{\omega}-x}^2dy\times\nonumber\\
& &\times e^{\alpha(\tilde{\omega}^2-x^2-y^2)}
[-\tilde{\omega}^4\nonumber\\
& &\qquad +2\tilde{\omega}^2(x^2+y^2)-(x^2-y^2)^2]\nonumber\\
& &+{{\pi\tau\Theta(2-\tilde{\omega})}\over{16\tilde{\omega}\overline{Z}}}
\int_{2-\tilde{\omega}}^2dx
\int_{4-x-\tilde{\omega}}^2dy\times\nonumber\\ 
& &\times e^{\alpha(\tilde{\omega}^2-x^2-y^2)} 
[-\tilde{\omega}^4\nonumber\\ 
& &\qquad +2\tilde{\omega}^2(x^2+y^2)-(x^2-y^2)^2]
\nonumber\\
& &+{{\pi\tau\Theta(2-\tilde{\omega})}\over{16\tilde{\omega}\overline{Z}}}
\int_0^{\tilde{\omega}}dr
\int_{\tilde{\omega}}^{4-\tilde{\omega}}ds  e^{\alpha[\tilde{\omega}^2-(r^2+s^2)/2]}
\times\nonumber\\
& &\times
[-\tilde{\omega}^4+\tilde{\omega}^2(r^2+s^2)-r^2s^2].\label{I1omega}
\end{eqnarray}

Last, but by no means least, we evaluate $\tilde{I}_3(\omega)$. This is easiest to do 
if we first use the trigonometric relations to rewrite Eq. (\ref{I3t}) in terms of 
$\cos[(y+s)t/\tau]$ and $\cos[(y-s)t/\tau]$.  Then, after Fourier 
transformation, we obtain,
\begin{eqnarray}
\tilde{I}_3(\omega)&=&{{\pi\tau}\over{16\overline{Z}}}\int_0^2dx
\int_0^2dy{{(1-y^2/4)}\over{y^2}}
\int_{|x-y|}^{x+y}{{ds}\over{s}}\times\nonumber\\ & &\times 
e^{\alpha(s^2-x^2-y^2)} 
\Bigl([(y+s)^2-x^2]^2[\delta(\tilde{\omega}+y-s)\nonumber\\ 
& &\qquad\qquad+\delta(\tilde{\omega}-y+s)] \nonumber\\
& & +[(y-s)^2-x^2]^2[\delta(\tilde{\omega}+y+s)\nonumber\\ 
& &\qquad\qquad+\delta(\tilde{\omega}-y-s)]\Bigr).\label{I3omegadef}
\end{eqnarray}
For $\omega\ge0$, the term containing $\delta(\tilde{\omega}+y+s)$ vanishes, so we are 
left with three terms, which we examine separately.  We denote them $\tilde{I}_{3n}$, 
where $n=1,2,3$, corresponding  to the order in which the remaining 
$\delta$-functions appear in Eq. (\ref{I3omegadef}).  The first integral, 
$\tilde{I}_{31}$, is subject to the constraints $|x-y|\le\tilde{\omega}+y\le x+y$.  
The second ($\tilde{\omega}+y\le x+y$) of these two constraints implies 
$\tilde{\omega}\le x$, which also implies $0\le\tilde{\omega}\le2$.  If $y<x$, the 
first constraint implies $y\ge(x-\tilde{\omega})/2$.  On the other hand, 
if $x>y$, there is no additional constraint on the integration region, other than 
$x\ge\tilde{\omega}$.  Thus, the combined integration region of this integral
is the interior of an 
irregular quadrangle with sides obeying $x=\tilde{\omega}$, $y=2$, $x=2$, 
and $y=(x-\tilde{\omega})/2$.  
Hence, we may now write $\tilde{I}_{31}(\omega)$ by inspection,
\begin{eqnarray}
\tilde{I}_{31}(\omega)&=&{{\pi\tau\Theta(2-\tilde{\omega})}
\over{16\overline{Z}}}
\int_{\tilde{\omega}}^2dx\int_{(x-\tilde{\omega})/2}^2dy
{{(1-y^2/4)}
\over{y^2(\tilde{\omega}+y)}}\times\nonumber\\
& &\times e^{\alpha(\tilde{\omega}^2+2\tilde{\omega}y-x^2)} 
[(2y+\tilde{\omega})^2-x^2]^2.\label{I31}
\end{eqnarray}

The integration regime of the second integral, $\tilde{I}_{32}(\omega)$ is 
subject to the constraints $|x-y|\le y-\tilde{\omega}\le x+y$.   The 
 second ``constraint'' makes no restrictions for $\tilde{\omega}\ge0$.  However, 
the first constraint can be rewritten as $\tilde{\omega}\le y-|x-y|$, and since $x$ 
and $y$ are both between 0 and 2, this then implies $0\le\tilde{\omega}\le 2$.  
For $x>y$, the first constraint further  requires $y\ge(\tilde{\omega}+x)/2$.  On the 
other hand, for $y>x$, the first constraint requires $x\ge\tilde{\omega}$.  Thus, 
the integration region is the interior of the irregular quadrangle with sides obeying 
$x=\tilde{\omega}$, $y=2$, $x=2$, and $y=(\tilde{\omega}+x)/2$.  We therefore write
\begin{eqnarray}
\tilde{I}_{32}(\omega)&=&{{\pi\tau\Theta(2-\tilde{\omega})}\over{16\overline{Z}}}
\int_{\tilde{\omega}}^2dx\int_{(\tilde{\omega}+x)/2}^2dy{{(1-y^2/4)}
\over{y^2(y-\tilde{\omega})}}\times\nonumber\\
& &\times e^{\alpha(\tilde{\omega}^2-2y\tilde{\omega}-x^2)} 
[(2y-\tilde\omega)^2-x^2]^2.\label{I32}
\end{eqnarray}

Finally, the integration region of $\tilde{I}_{33}(\omega)$ is subject to the 
constraints $|x-y|\le\tilde{\omega}-y\le x+y$.   The second constraint implies 
that $\tilde{\omega}$ can be as large as 6, and also that $y\ge(\tilde{\omega}-x)/2$.  
For $x>y$, the first constraint  implies $x\le\tilde{\omega}$, whereas for $y>x$, it 
implies $y\le(\tilde{\omega}+x)/2$.  For $0\le\tilde{\omega}\le2$, these constraints 
restrict the integration region to the interior of the triangle with sides obeying 
$y=(\tilde{\omega}-x)/2$, $y=(\tilde{\omega}+x)/2$, and $x=\tilde{\omega}$.  
For $4\le\tilde{\omega}\le6$, the integration region is the interior of the
triangle
with sides obeying $y=2$, $x=2$, and $y=(\tilde{\omega}-x)/2$.  In the intermediate 
regime $2\le\tilde{\omega}\le4$, the integration region is the interior of the 
irregular quadrangle with sides obeying  $y=(\tilde{\omega}-x)/2$, 
$y=(\tilde{\omega}+x)/2$, $y=2$, and $x=2$.  We break this integration region up 
into two parts.  One of these  parts is the interior of the isosceles triangle 
with sides obeying $y=(\tilde{\omega}-x)/2$, $y=(\tilde{\omega}+x)/2$, and 
$x=4-\tilde{\omega}$. The second region is the interior of the irregular
quadrangle 
with sides obeying  $x=4-\tilde{\omega}$, $y=2$, $x=2$, and $y=(\tilde{\omega}-x)/2$.   
Altogether, we write $\tilde{I}_{33}(\omega)$ as 
\begin{eqnarray}
\tilde{I}_{33}(\omega)&=&{{\pi\tau\Theta(2-\tilde{\omega})}\over{16\overline{Z}}}
\int_0^{\tilde{\omega}}dx\int_{(\tilde{\omega}-x)/2}^{(\tilde{\omega}+x)/2}dy 
f(x,y,\tilde{\omega})\nonumber\\
& &+{{\pi\tau\Theta(\tilde{\omega}-2)\Theta(4-\tilde{\omega})}\over{16\overline{Z}}}
\times
\nonumber\\ 
& &\times\Biggl[\int_0^{4-\tilde{\omega}}dx
\int_{(\tilde{\omega}-x)/2}^{(\tilde{\omega}+x)/2}
dy \nonumber\\ 
& &\qquad
 +\int_{4-\tilde{\omega}}^2dx\int_{(\tilde{\omega}-x)/2}^2dy\Biggr]
f(x,y,\tilde{\omega})
\nonumber\\
& &+{{\pi\tau\Theta(\tilde{\omega}-4)\Theta(6-\tilde{\omega})}\over{16\overline{Z}}}
\times
\nonumber\\
 & &\qquad\times\int_{\tilde{\omega}-4}^2dx\int_{(\tilde{\omega}-x)/2}^2dy 
f(x,y,\tilde{\omega}),\label{I3omega}\\
\noalign{\rm where}
f(x,y,\tilde{\omega})&=&{{(1-y^2/4)}\over{y^2(\tilde{\omega}-y)}}
e^{\alpha(\tilde{\omega}^2
-2\tilde{\omega}y-x^2)}\times\nonumber\\ 
& &\qquad\times[(2y-\tilde{\omega})^2-x^2]^2.\label{I3integrand}
\end{eqnarray}

In Figs. 16 - 20, we present our results for the $\delta\tilde{{\cal C}}_{ij}(\omega)$, 
plotted as functions of $\omega\tau$.  In the figure labels, we drop the
tildes for clarity.  Note that these functions are rigorously zero 
for $\omega\tau > 6$, but they are so small for $\omega\tau>4$ that they are
indistinguishable from zero for $\omega\tau>4.1$ in plots with $0\le\omega\tau\le6$. In each case, $\delta\tilde{{\cal C}}_{12}(\omega)$ are the 
solid curves, $\delta\tilde{{\cal C}}_{24}(\omega)$ are the dotted curves, and 
$\delta\tilde{{\cal C}}_{22}(\omega)$ are the dashed curves.   At infinite
temperature, these are obtained from  our exact formulae for the $f_i(s)$,
$g_i(s)$ and $h_3(s)$ listed in the Appendix
by simply letting $s\rightarrow\omega\tau$, and multiplying the overall
results by $\pi$.   For example,
$\delta\tilde{\cal C}_{12}(\omega)= -\pi[f_1(\omega\tau)+g_1(\omega\tau)]$.
The exact expressions for the three $\delta\tilde{\cal C}_{ij}(\omega)$ are plotted in Fig. 16.   As a check on our computations, we also obtained 
these results by numerically evaluating the $\delta 
{\cal C}_{ij}(t)$ from the double integrals in the infinite temperature limit, and the results were found to agree to 
within three significant figures.  

\begin{figure}
\vspace{0.3cm}
\epsfxsize=8cm
\centerline{\epsffile{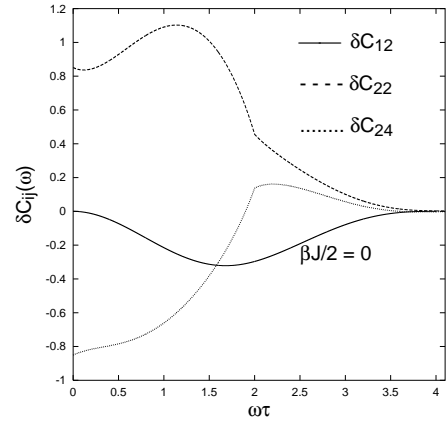}}
\vskip0.3cm
\caption{Plots of the  exact Fourier transforms $\delta\tilde{C}_{12}(\omega)$ (solid), 
$\delta\tilde{C}_{22}(\omega)$ (dashed), and $\delta\tilde{C}_{24}(\omega)$ (dotted) 
of $C_{ij}(t)-{\rm lim}_{t\rightarrow\infty}C_{ij}(t)$, as functions of $\omega\tau$, 
in the infinite temperature limit $\alpha=0$.}\label{fig16}
\end{figure}

\begin{figure}
\vspace{0.3cm}
\epsfxsize=8cm
\centerline{\epsffile{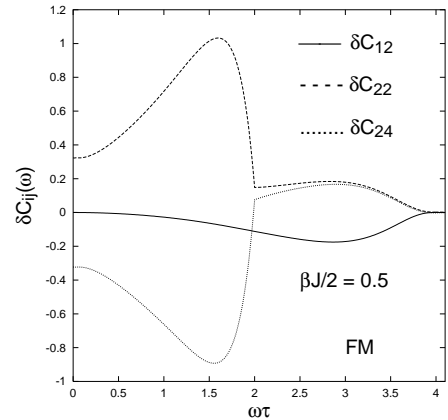}}
\vskip0.3cm
\caption{Plots of the  Fourier transforms $\delta\tilde{C}_{12}(\omega)$ (solid), 
$\delta\tilde{C}_{22}(\omega)$ (dashed), and $\delta\tilde{C}_{24}(\omega)$ (dotted) 
of $C_{ij}(t)-{\rm lim}_{t\rightarrow\infty}C_{ij}(t)$, as functions of
$\omega\tau$, for the FM case $\alpha=0.5$}\label{fig17}
\end{figure}

\begin{figure}
\vspace{0.3cm}
\epsfxsize=8cm
\centerline{\epsffile{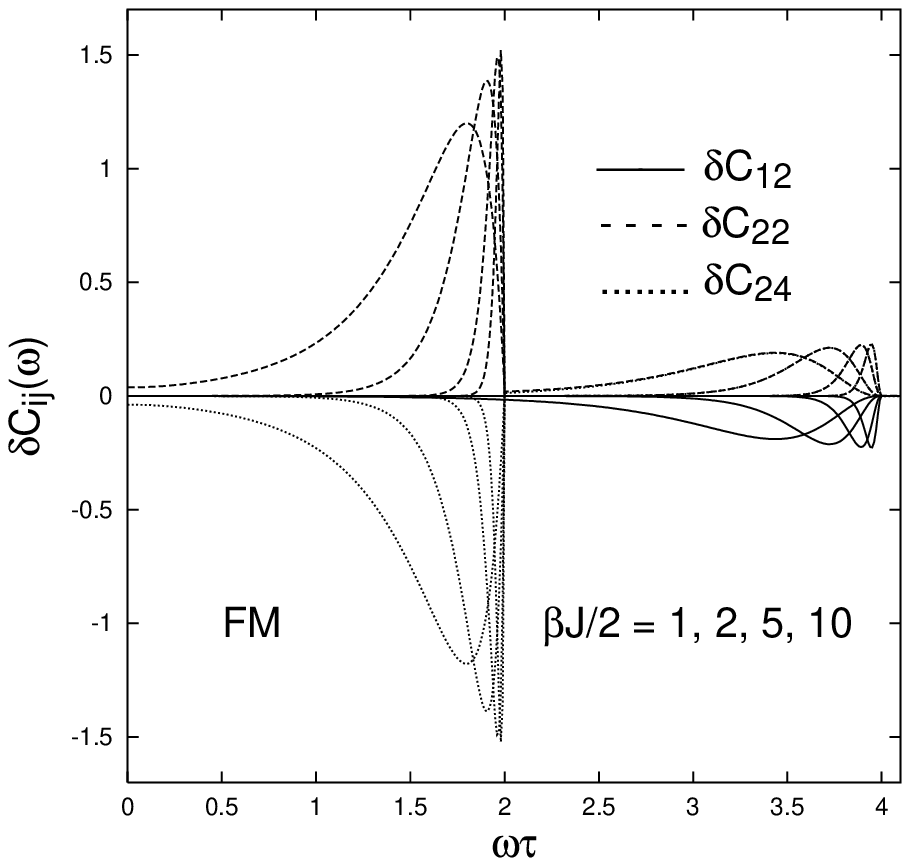}}
\vskip0.3cm
\caption{Plots of the  Fourier transforms $\delta\tilde{C}_{12}(\omega)$ (solid), 
$\delta\tilde{C}_{22}(\omega)$ (dashed), and $\delta\tilde{C}_{24}(\omega)$ (dotted) 
of $C_{ij}(t)-{\rm lim}_{t\rightarrow\infty}C_{ij}(t)$, as functions of
$\omega\tau$, for the FM cases $\alpha=1, 2, 5,$ and 10.   The peaks
developing successively near to  $\omega\tau=2, 4$ with decreasing $T$ are magnons. }
\label{fig18}
\end{figure}

In this and in subsequent figures, we also checked 
the accuracy of our analytic formulae by the zero-time sum rule, the 
$\int_0^{\infty}\delta\tilde{{\cal C}}_{ij}(\omega) d\omega/\pi=\delta {\cal C}_{ij}(0)$.  It is 
seen that the functions all are continuous, and approach zero at large $\omega$.  
$\delta\tilde{{\cal C}}_{12}(\omega)\le0$, and $\delta\tilde{{\cal C}}_{22}(\omega)\ge0$, but 
$\delta\tilde{{\cal C}}_{24}(\omega)$ has regions of both signs.  
In Figs. 17 and 18, we present the data for the FM case, with $\alpha= 0.5$ in
Fig. 17, and $\alpha= 1, 2,
 5$, and 10 in Fig. 18.  As the temperature is lowered, the peak in 
$\delta\tilde{{\cal C}}_{22}(\omega)$ moves to higher $\omega$ values.  Moreover,  
$\delta\tilde{{\cal C}}_{22}(\omega)$ and $\delta\tilde{{\cal C}}_{24}(\omega)$ approach each 
other in the region $2\le\omega\tau\le4$, but approach the opposite of each other in 
the regime $0\le\omega\tau\le2$. Thus, in the regime $2\le\omega\tau\le4$ of
Fig. 18, these 
dashed and dotted curves combine to give a curve that appears 
to be dash-dotted.  In addition, as the temperature is lowered, the curves all 
develop into sharp peaks at $\omega\tau\approx2$ and 4, which are asymmetric, 
dropping rapidly to zero 
at $\omega\tau=2$ and 4, but having longer tails at lower $\omega\tau$ values. 
 More precisely, $\delta\tilde{{\cal C}}_{12}(\omega)$ becomes a single peak 
at 
$\omega\tau\approx 4$, whereas the other two, $\delta\tilde{{\cal C}}_{22}(\omega)$ and 
$\delta\tilde{{\cal C}}_{24}(\omega)$  have identical peaks at $\omega\tau\approx 4$,
 which are opposite to that of $\delta\tilde{{\cal C}}_{12}(\omega)$.  But, they also 
have larger and sharper peaks at $\omega\tau\approx2$, which are opposite in sign to 
each other.  These peaks at $\omega\tau\approx 2, 4$ arise from magnons.  Hence, Fig. 18 provides a simple explanation of the detail of 
${\cal C}_{ij}(t)$ for $\alpha=10$ shown in Fig. 8.

\begin{figure}
\vspace{0.3cm}
\epsfxsize=8cm
\centerline{\epsffile{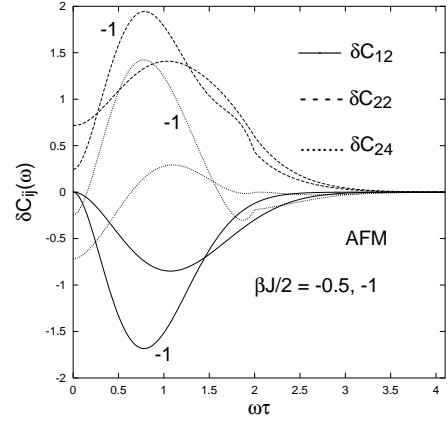}}
\vskip0.3cm
\caption{Plots of the  Fourier transforms $\delta\tilde{C}_{12}(\omega)$ (solid), 
$\delta\tilde{C}_{22}(\omega)$ (dashed), and $\delta\tilde{C}_{24}(\omega)$ (dotted) 
of $C_{ij}(t)-{\rm lim}_{t\rightarrow\infty}C_{ij}(t)$, as functions of
$\omega\tau$, for the AFM cases $\alpha=-0.5$ and -1.  The bumps developing for
$\alpha=-1$ at $\omega\tau\approx2$ are magnons.}
\label{fig19}
\end{figure}

\begin{figure}
\vspace{0.3cm}
\epsfxsize=8cm
\centerline{\epsffile{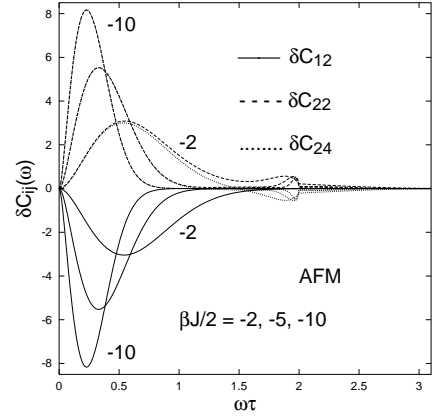}}
\vskip0.3cm
\caption{Plots of the  Fourier transforms $\delta\tilde{C}_{12}(\omega)$ (solid), 
$\delta\tilde{C}_{22}(\omega)$ (dashed), and $\delta\tilde{C}_{24}(\omega)$ (dotted) 
of $C_{ij}(t)-{\rm lim}_{t\rightarrow\infty}C_{ij}(t)$, as functions of
$\omega\tau$, for the AFM cases $\alpha=-2, -5,$ and -10.   The peaks for
$\omega\tau\approx 2$ that
successively sharpen  with decreasing $T$ are magnons.}\label{fig20}
\end{figure}

Curves for the AFM case are shown in Fig. 19 and 20.  In Fig. 19, we display
the  results for $\alpha=-0.5$ and -1 together, and in Fig. 20, the results
for $\alpha= -2, -5,$ and -10 are shown.  As the temperature is lowered, 
the peak in $\delta\tilde{{\cal C}}_{22}(\omega)$ moves to lower frequency, resulting in 
the slowing down seen in the real time curves.  It develops into two 
peaks, a large one at low $\omega\tau$, and a small one at
$\omega\tau\approx2$, which is an antiferromagnetic magnon.  Note that the
magnon width sharpens as $T$ is lowered.  
Surprisingly, $\delta\tilde{{\cal C}}_{24}(\omega)$ changes dramatically, mostly changing 
sign near $\omega\tau\approx 2$, but the small $\omega$ behavior increases to join the small $\omega$ behavior 
of $\delta\tilde{{\cal C}}_{22}(\omega)$, and the opposite of it in the region 
$\omega\tau\approx2$. These peaks at $\omega\tau\approx 2$ arise from
antiferromagnetic magnons.  In addition, $\delta\tilde{{\cal C}}_{12}(\omega)$ stays negative, and 
develops into a negative peak at low $\omega$ which is opposite 
to that of $\delta\tilde{{\cal C}}_{22}(\omega)$ and $\delta\tilde{{\cal
C}}_{24}(\omega)$.  

Finally, in order to elucidate the nature of the slowing down as $T$ is
lowered, in Fig. 21 we plotted our AFM results for $\alpha = -0.5,-1,-2,-5$, 
and
-10 together, as a function of the scaled frequency $\omega\tau|\beta
J/2|^{1/2}$.  We note that the position of the low frequency peak does indeed
scale, but since the magnon appears at a fixed frequency, it does not scale.
The magnon is distinctly visible for  $\alpha=-2,-5,$ and $-10$ in this
figure, as noted by the arrows.

\begin{figure}
\vspace{0.3cm}
\epsfxsize=8cm
\centerline{\epsffile{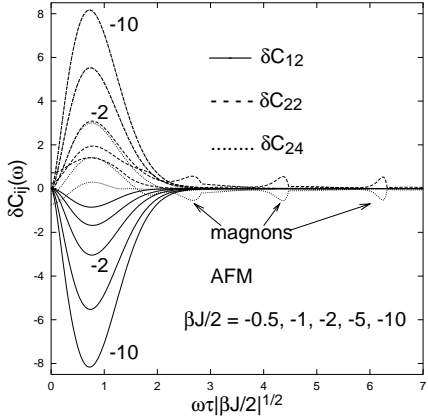}}
\vskip0.3cm
\caption{Plots of the  Fourier transforms $\delta\tilde{C}_{12}(\omega)$ (solid), 
$\delta\tilde{C}_{22}(\omega)$ (dashed), and $\delta\tilde{C}_{24}(\omega)$ (dotted) 
of $C_{ij}(t)-{\rm lim}_{t\rightarrow\infty}C_{ij}(t)$, as functions of the
scaled frequency $\omega\tau|\beta J/2|^{1/2}$,
 for the AFM cases $\alpha=-0.5, -1, -2, -5,$ and -10.   The magnon  peaks
for  $\omega\tau\approx 2$
do not scale, and are marked with arrows.}\label{fig21}
\end{figure}

\section{Discussion and Conclusions}

In this paper we have presented the exact solution for the thermal equilibrium dynamics of four classical Heisenberg spins on
a ring.  For this system, there are three relevant time correlation functions,
which are the auto- (${\cal C}_{22}$),  near-neighbor (${\cal C}_{12}$), and
the next-nearest-neighbor (${\cal C}_{24}$) 
correlation functions, respectively.  Using our
results, we wrote exact expressions for these three functions in terms of triple integrals.  At infinite temperature,
we reduced these triple integrals to single integrals.  We also
obtained analytic expressions for the long-time, infinite temperature behavior
of the three correlation functions. We found that the near-neighbor
correlation function ${\cal C}_{12}(t)$ is strikingly different from the auto- and
next-nearest-neighbor correlation functions ${\cal C}_{22}(t)$ and ${\cal C}_{24}(t)$.  Not only does it approach a
different finite value asymptotically, but it also oscillates with twice the 
frequency,
and the oscillations  decay in amplitude much more rapidly.   Although the
long-time asymptotic values of these functions vary with temperature, the
near-neighbor correlation function similarly differs from the other two
functions for all temperatures.

In addition,
we were able to obtain the Fourier transforms of the deviations of the
correlation functions from their infinite time aymptotic limits in terms of double
integrals.  At infinite temperature, exact analytic forms for these Fourier
transforms were obtained.  As the temperature is lowered, peaks in the Fourier
transforms appear at $\omega\tau=2$ for  antiferromagnetic coupling, and at
$\omega\tau=2,4$ for ferromagnetic coupling.  These peaks sharpen up as the
temperature is lowered.  Although the origin of these peaks  is purely
classical, they correspond
precisely to magnons, which are usually thought of as quantum mechanical in
origin.  Here the magnons arise from standing waves, such as those on a violin
string. These standing waves arise from the periodicity of the ring, as a
combination of traveling waves moving both clockwise and counterclockwise in direction.  For the
antiferromagnetic case, neighboring spins are opposite in direction at low
temperatures, so that two full wavelengths fit into the ring.  For the
ferromagnetic case, one can have either two or four full wavelengths in the
ring, the latter corresponding to every spin pointing in the same direction.

In the quantum mechanical analogue, the magnon  energies are
$E_n=2J[1-\cos(k_na)]$, where $k_n=2n\pi/L$ and $L=4a$. We thus get $E_0=0$,
$E_1=2J$, and $E_2=4J$.  These latter two values correspond precisely to $\omega\tau=2,4$.

 In addition, for antiferromagnetic coupling, a second, much larger
peak in the Fourier transform functions appears at increasing lower frequency
as the temperature is lowered.  This second peak is found to scale with
$\omega/T^{1/2}$, corresponding to the $tT^{1/2}$ scaling of the low
temperature  antiferromagnetic time correlation functions.

It is interesting to compare the long-time asymptotic results of the
correlation functions at infinite temperature with the known results for
other Heisenberg rings.  For a dimer, with two Heisenberg spins interacting
via Eq. (1) with ${\bf S}_3={\bf S}_1$,  the
infinite temperature limit of the auto- and near-neighbor correlation
functions were given for all $t$, \cite{LL}
\begin{eqnarray}
\lim_{T\rightarrow\infty}{\cal
C}_{11}(t)&=&{{1}\over{2}}-{{1+2\cos(2t^*)}\over{2t^{*2}}}\nonumber\\
& &+{{3\sin(3t^*)}\over{2t^{*3}}}-{{3[1-\cos(2t^*)]}\over{4t^{*4}}}\\
&=&1-\lim_{T\rightarrow\infty}{\cal
C}_{12}(t),\\
\end{eqnarray}
where we have used the conservation law, 
\begin{equation}
{\cal C}_{11}(t)+{\cal
C}_{12}(t)=\langle S^2\rangle/2,
\end{equation}
and $\lim_{T\rightarrow\infty}\langle S^2\rangle/2=1$ for the dimer.

For the case of three spins on a ring, it is not so trivial, but it
is still much easier to evaluate the correlation functions at large times
and infinite temperature than for the four-spin ring.  The infinite
temperature aymptotic limit result  for the 
autocorrelation function ${\cal
C}_{11}(t)$ was quoted previously, \cite{LL} and its complete derivation
for all temperatures was given. \cite{LBC} 
In this case, reduction to quadrature is rather simple, and one obtains a
result for the three-spin ring $\delta{\cal C}_{11}(t)$ analogous to
 Eq. (\ref{f1g1}).  At infinite temperature, \cite{LBC}
\begin{eqnarray}
\lim_{T\rightarrow\infty}\delta{\cal C}_{11}(t)&=&\int_0^1ds f_4(s)\cos(st^*)\nonumber\\
& &+\int_1^3ds g_4(s)\cos(st^*),
\end{eqnarray}
where $f_4(s)$, $g_4(s)$, and their values and relevant derivatives at the integration endpoints
 are given in the Appendix.
Thus,  for the three-spin ring,  the leading terms for long times were shown
 to be, \cite{LBC}
\begin{eqnarray}
\lim_{{T\rightarrow\infty}\atop{t\gg\tau}}{\cal
  C}_{11}(t)&=&{1\over{3}}+\delta_3-{{[\sin(t^*)+\sin(3t^*)]}\over{t^{*3}}},\\
\noalign{\rm where}\nonumber\\
\delta_3&=&{{9}\over{40}}\ln 3-{1\over{10}}\approx 0.147188.
\end{eqnarray}
  In analogy with
Eq. (\ref{s2}),  it is then easy to see that 
\begin{eqnarray}
2{\cal C}_{12}(t)+{\cal C}_{11}(t)&=&\langle S^2\rangle/3,\\
\noalign{\rm and hence that}\nonumber\\
\lim_{{T\rightarrow\infty}\atop{t\gg\tau}}{\cal
  C}_{12}(t)&=&{1\over{3}}-{{\delta_3}\over{2}}+{{[\sin(t^*)+\sin(3t^*)]}\over{2t^{*3}}},
\end{eqnarray}
since $\lim_{T\rightarrow\infty}\langle S^2\rangle/3=1$ for the three-spin 
ring.

For $N\ge3$ classical Heisenberg spins on a ring, it is straightforward to obtain
the conservation equation for the correlation functions,
\begin{eqnarray}
{\cal C}_{11}(t)+2\sum_{n=2}^{(N+1)/2}{\cal C}_{1n}(t)&=&\langle S^2\rangle/N,
\\
\noalign{\rm and}\nonumber\\
{\cal C}_{11}(t)+2\sum_{n=2}^{N/2}{\cal C}_{1n}(t)+{\cal
C}_{1,(N+2)/2}(t)&=&\langle S^2\rangle/N,
\end{eqnarray}
for $N$ odd and even, respectively.

We note  that for $1\le N\le 4$ at infinite temperature, $\langle
S^2\rangle/N=1$.   More generally, at infinite temperature, the evaluation of 
$\langle S^2\rangle$ for an $N$-spin ring
maps onto that of the mean square displacement of a chain of length $N$
during a random walk in three dimensions,
and hence rigorously  $\lim_{T\rightarrow\infty}\langle
S^2\rangle/N=1\>\> \forall\> N\ge1$.  We
then note that the long-time behavior of ${\cal C}_{11}(t)$ at infinite
temperature arises primarily from the discontinuities in the derivatives of the
functions $f_i(s)$, $g_i(s)$, $h_i(s)$, etc. at the  endpoints of the
integration intervals.  For $N=2,4$, the functions are finite
and continuous over the integration regions, but at least one of their first
derivatives is discontinuous at one or more of the integration endpoints.    In addition, for the  particular odd-spin ring with $N=3$,
both the functions and their first derivatives are continous at the
integration endpoints, but the second derivatives are discontinuous at the
endpoints.  The mathematical forms of the functions $f_i(s)$, $g_i(s)$,
$h_i(s)$, etc. and their various derivatives at the integration endpoints become
increasingly complicated with increasing $N$ for $N\le4$.

 It would
be 
interesting to find out whether this ``pattern'' of matching the functions and
their derivatives at the integration endpoints might be maintained for much
larger $N$ values.  Thus, at least two possible scenarios that might develop from attempting to
generalize our results to much larger $N$ values arise.  Regardless of whether
$N$ is even or odd,
 ${\cal C}_{11}(t)$
might behave for $t^{*}\gg1$ as 
\begin{eqnarray}
{\cal
 C}_{11}(t)&\rightarrow&\left\{ \begin{array}{l}
 {1\over{N}}+\delta_N+
\sum_{n=0}^{N}a_n\cos(nt^*)/t^{*2},\\
\\
{1\over{N}}+\delta_N+\sum_{n=1}^Nb_n\sin(nt^*)/t^{*},
\end{array}
\right.
\label{Nevenodd}
\end{eqnarray}
which would occur if the functions were finite and continuous but with one or
more discontinuous first derivatives, as for $N=2,4$, or if they were finite but
discontinous at one or more of the integration endpoints, respectively.  Of course, behavior
such as for $N=3$ could also be obtained for higher $N$ values, as well, as
well as more complicated scenarios.   We expect that $\lim_{N\rightarrow\infty}\delta_N=0$, and 
  that $\delta_N$ decreases to zero faster than $1/N$.  We note that for 
$N=2$, we have $\delta_2=0$. 
  
Numerical  simulation data for the $N=4,6,8,10,50$ autocorrelation functions were presented for
$0\le t^*\le10$ by M{\"uller}. \cite{Mueller}    Those results show that the oscillations in ${\cal
C}_{11}(t)$ for $N\ge4$ are easily discernible out to $t^*=10$.  Hence, it appears unlikely
that the autocorrelation functions would approach their asymptotic limits more
rapidly than $1/t^{*2}$. The high accuracy of the numerical results is exemplified by the excellent agreement between the result for
$N=4$ and our own exact analytical results, \cite{Mueller} as shown in Fig. 3b.

The main difference between rings with even
and odd numbers $N$ of spins lies in the long-time behaviors
of the two-spin time correlation functions.  For both $N$ even and odd, the 
conservation law  requires at least one  of the
${\cal C}_{1n}$ for $n\ne1$ to compensate for the leading long-time behavior
of ${\cal C}_{11}(t)$.  For odd $N$, we anticipate that all of the correlation
functions  will fall off with oscillatory corrections that have an amplitude
of order $(t^{*})^{-m}$ at large times, where $m$ is likely to be a small natural number. 
We then raise the question for even $N$,  as to which,
if any,  of the
correlation functions approaches its asymptotic limit more rapidly than the autocorrelation function does
for long times.
 Based upon the $N=2,4$ examples, the answer to this question might 
depend upon whether
$N/2$ is even or odd.    

Finally, we turn to the unresolved question, vigorously debated in the
literature for over a decade,\cite{Mueller,Landau,Mueller1,Bonfim} of whether
the large-$N$ limit of the two-spin correlation functions for rings of
classical Heisenberg spins, based on the dynamics of Eq. (\ref{Sidot}), will
decay to zero with 
 the leading behavior $t^{-1/2}$ for
long times.  This asymptotic  behavior can easily be derived for an infinite
linear chain of classical spins whose dynamics are governed by the following
discretized version of a continuous spin hydrodynamics,
\begin{equation}
{{d{\bf S}_{l}}\over{dt}}=\gamma({\bf S}_{l+1}+
{\bf S}_{l-1}-2{\bf S}_l).\label{diffuse}
\end{equation}
The parameter $\gamma=D/a^2$, where $a$ is the lattice constant, and $D$ is a
spin diffusion coefficient.  The final result for the vector component $\mu$
of the two-spin correlation
function for a pair of spins $n$ lattice sites apart coincides with the
probability distribution for a one-dimensional continuous time random walk from the origin, \cite{Montroll}
\begin{equation}
\langle S^{\mu}_{l}(0)S^{\mu}_{l+n}(t)\rangle=\exp(-t^*)I_n(t^*),
\label{Monty}
\end{equation}
where now $t^*=2\gamma t$ and $I_n(z)$ is a modified Bessel function.  For fixed $n$, the leading
behavior for large $t$ is indeed proportional to $(t^*)^{-1/2}$.

For the four-spin ring with diffusive dynamics,
Eq. (\ref{diffuse}), one finds at
infinite temperature,
\begin{eqnarray}
{\cal C}_{22}(t)&=&[1+\exp(-t^*)]^2/4,\label{C22diffuse}\\
{\cal C}_{24}(t)&=&[1-\exp(-t^*)]^2/4,\\
\noalign{\rm and}\nonumber\\
{\cal C}_{12}(t)&=&[1-\exp(-2t^*)]/4,\\
\noalign{\rm but}\nonumber\\
\langle {\bf S}_i^2(t)\rangle&=&[1+\exp(-2t^*)]^2/4\label{magspin}
\end{eqnarray}
for $i=1,\ldots,4$.  Although the three correlation functions are different,
satisfying the conservation law, Eq. (\ref{s2}), and ${\cal C}_{12}(t)$
correctly reaches its asymptotic limit faster than do either of the other two,
they all go exponentially to the same asymptotic value $1/4$, and do not exhibit any oscillations.  In
addition,  the rms value of each individual spin magnitude
decays from its assumed initial value of unity to 1/2.

Generalizing this  treatment to a closed finite ring with $N$ sites, as
derived using Eq. (\ref{diffuse}), leads to the following results. The corresponding time correlation functions all approach the same non-zero limit
$1/N$ for long times, with correction terms that rise or decay as
$\exp(-\mu_Nt^*$), where $\mu_N=1-\cos(2\pi/N)$.  In addition, each rms individual spin magnitude decays exponentially
from its initial value of unity to $1/N^{1/2}$.  These features are distinctly different from the exact results we have obtained for the
four-spin ring, and others have obtained for the dimer and three-spin 
ring, \cite{LL,LBC}
based upon Heisenberg dynamics, Eq. (\ref{Sidot}). However, the diffusive
approximation does suggest that for $N$ divisible by 4, ${\cal
C}_{1,1+N/4}(t)$ should approach its asymptotic limit faster than the other
correlation functions, as occurs in the exact treatment for $N=4$.  It would be
interesting to see if qualitatively similar features are obtained for higher
$N$ values using Heisenberg dynamics. 
  
It is noteworthy that as
long as $t$ is sufficiently small compared to $N^2/(2\gamma)$, the numerical
results of the treatment for the $N$-spin ring based upon Eq. (\ref{diffuse})
are virtually indistinguishable from the result, Eq. (\ref{Monty}), for the
infinite chain.  This can be expected, since the rms distance achieved by the
corresponding random walker remains small compared to the circumference of the
ring, hence for all intents and purposes the behavior should be the same as
that of an infinite linear chain.

Now our main point is that  diffusive spin dynamics based upon
Eq. (\ref{diffuse}), while conserving the components of the total spin
vector, {\it does not} preserve the length of the individual spin vectors.
However, this
property, $|{\bf S}_i(t)|=1$ for $i=1,\ldots,N$,  is of course maintained at
all times by  Heisenberg dynamics, Eq. (\ref{Sidot}).
 That is, {\it the spin dynamics based upon
Eq. (\ref{diffuse}) are fundamentally different in character from those based 
upon
Eq. (\ref{Sidot})}. Not surprisingly, the present exact results at infinite temperature,
Eqs. (\ref{f1g1}) - (\ref{f3g3h3}), displayed in Fig. 3a, along with the analytic expansions for long times, Eqs. (\ref{C12longtime}) - (\ref{C24longtime}),   of the two-spin
correlation functions for $N = 4$ spins based upon Heisenberg dynamics,
Eq. (\ref{Sidot}), differ both quantitatively and qualitatively from the
results, Eqs. (\ref{C22diffuse}) - (\ref{magspin}), that one obtains using diffusive dynamics from the four-spin version
of Eq. (\ref{diffuse}).    These fundamental differences 
only underscore the fact that the broader question, of whether the time
correlation functions derived from the Heisenberg spin dynamics,
Eq. (\ref{Sidot}), will show a $t^{-1/2}$ long-time approach to their
asymptotic limits, as does Eq. (\ref{Monty}), 
is acute and deserving of greater attention.

\section{acknowledgments}

The authors would like to thank T. Bandos, D. Mentrup, K. Scharnberg, J. Schnack, 
and C. Schr{\"o}der for very useful discussions.  One of us (RAK) would like to thank 
the Ames Laboratory and the Max-Planck-Institut f{\"u}r Physik komplexer Probleme for 
their hospitality.  The Ames Laboratory is operated by Iowa State University under 
Contract No. W-7405-Eng-82.

\section{Appendix}
Here we list the functions $f_i(s)$, $g_i(s)$, and $h_i(s)$ for
the correlation functions at infinite temperature.  We find,
\begin{eqnarray}
f_1(s)&=&{1\over{144}}s^2(s-2)^2(s+4)\nonumber\\
& & - {{s^3}\over{2880}}(5s^2+48s-150),\\
g_1(s)&=&{{(4-s)^3}\over{2880s}}(5s^3+12s^2-6s-8),\\
f_2(s)&=&{{(4-s^2)}\over{480s^2}}\biggl[12s+11s^3+s^5\ln\Bigl({{s^2}\over{4-s^2}}
\Bigr)\nonumber\\
& &-(12-10s^2+{{15}\over{4}}s^4)\ln\Bigl({{2+s}\over{2-s}}\Bigr)\biggr],\\ 
f_3(s)&=& a_1(s)+a_2(s)(2-s)\ln{2}\nonumber\\
& &+{{s^3(40-s^2)}\over{480}}\ln\Bigl({{s}\over{s+2}}\Bigr)+\nonumber\\
& &+a_3(s)(2-s)^2\ln(4-s^2),\\
\noalign{\rm where}\nonumber\\
a_1(s)&=&{1\over{115200}}\Bigl(61264-28800s+19160s^2\nonumber\\
& &-4160s^3+45s^4+188s^5\Bigr),\\
a_2(s)&=&-{1\over{1920s^2}}\Bigl(96+48s+608s^2\nonumber\\
& & +304s^3-318s^4-23s^5-4s^6\Bigr),\\
\noalign{\rm and}\nonumber\\
a_3(s)&=&{1\over{320s^2}}(4+4s+13s^2+12s^3),\\
g_3(s)&=&b_1(s)+b_2(s)\ln{2}+{{s^3(s^2+32)}\over{480}}\ln{s}\nonumber\\
& &+b_3(s)\ln(s-2)+b_4(s)\ln(s+2),\\
\noalign{\rm where}\nonumber\\
b_1(s)&=&{1\over{230400s}}\Bigl(-17280 -28496s+60000s^2\nonumber\\
& &+21800s^3-5600s^4-1485s^5+188s^6\Bigr),\\
b_2(s)&=&-{{(s+2)}\over{3840s^2}}\Bigl(-96+48s-608s^2\nonumber\\
& &+304s^3+318s^4-23s^5+4s^6\Bigr),\\
b_3(s)&=&-{{(s-2)^2}\over{3840s^2}}\Bigl(96+96s+140s^2\nonumber\\
& &\qquad +116s^3+s^4+4s^5\Bigr),\\
\noalign{\rm and}\nonumber\\
b_4(s)&=&-{1\over{3840s^2}}(s-2)(s+2)^4(6-9s+4s^2),\\
\noalign{\rm and finally,}\nonumber\\
h_3(s)&=&{{(s-2)}\over{3840}}\biggl[s^3(23+4s)
\ln\Bigl({{2(s-2)}\over{s+2}}\Bigr)\nonumber\\
& &+2(92+46s-57s^2)\ln\Bigl({{s-2}\over{4}}\Bigr)\nonumber\\
& &-10(8+4s-3s^2)
\ln\Bigl({8\over{s+2}}\Bigr)\nonumber\\
& &+{{48(s+2)}\over{s^2}}
\ln\Bigl({{(s-2)^2}\over{2(s+2)}}\Bigr)\biggr]+c_1(s),\\
\noalign{\rm where}\nonumber\\
c_1(s)&=&{{(s-6)}\over{230400s}}\Bigl(2880+16152s-22668s^2\nonumber\\
& &+6242s^3+267s^4-188s^5\Bigr).\\
\noalign{\rm  For the three-spin ring,}\nonumber\\
f_4(s)&=&{{s^2(5-s^2)}\over{15}}\\
\noalign{\rm and}\nonumber\\
g_4(s)&=&{{(s-3)^2}\over{120s}}(-3-2s+9s^2+4s^3).
\end{eqnarray}

To obtain the leading long-time behavior of the
$\lim_{T\rightarrow\infty}{\cal C}_{ij}(t)$, we
require the functions $f_i(s),g_i(s)$, and $h_i(s)$ and their
derivatives at the integration endpoints $s = 0, 2, 4, 6$ (0, 1, 3 for the
three-spin ring).  The  relevant quantities  are
\begin{eqnarray}
f_1(0)&=&f_1^{'}(0)=g_1(4)=g_1^{'}(4)=g_1^{''}(4)=0,\\
f_1(2)&=&g_1(2)={{17}\over{180}},\\
f_1^{'}(2)&=&g_1^{'}(2)=-{{17}\over{360}},\\
f_1^{''}(0)&=&{2\over{9}},\\
f_1^{''}(2)&=&g_1^{''}(2)=-{{43}\over{360}},\\
f_1^{'''}(0)&=&-{3\over{16}},\\
f_1^{'''}(2)&=&g_1^{'''}(2)={{21}\over{80}},\\
g_1^{'''}(4)&=&-{1\over{4}},\\
f_2(0)&=&f_2(2)=h_3(6)=h_3^{'}(6)=0,\\
f_2^{'}(0)&=&{1\over{6}},\\
f_2^{'}(2)&=&-{7\over{30}}+{2\over{15}}\ln{2},\\
f_3(0)&=&{{3739}\over{7200}}-{{43}\over{120}}\ln{2},\\
f_3(2)&=&g_3(2)={7\over{15}}-{3\over{5}}\ln{2},\\
f_3^{'}(0)&=&-{1\over{4}},\\
f_3^{'}(2)&=&{{11}\over{40}}-{{29}\over{30}}\ln{2},\\
g_3^{'}(2)&=&{{21}\over{40}}-{{29}\over{30}}\ln{2},\\
g_3(4)&=&h_3(4)={{69}\over{1600}}+{{177}\over{80}}\ln{2}-{{459}\over{320}}\ln{3},\\
\noalign{\rm and}\nonumber\\
g_3^{'}(4)&=&h_3^{'}(4)={5\over{36}}+{{57}\over{20}}\ln{2}-{{1233}\over{320}}\ln{3}.\\
\noalign{\rm For the three-spin ring, we require}\nonumber\\
f_4(0)&=&f_4^{'}(0)=g_4(3)=g_4^{'}(3)=0,\\
f_4^{'}(1)&=&g_4^{'}(1)={2\over{5}},\\
f_4^{''}(1)&=&-{{2}\over{15}},\\
g_4^{''}(1)&=&-{{17}\over{15}},\\
\noalign{\rm and}\nonumber\\
g_4^{''}(3)&=&1.
\end{eqnarray}


\begin{references}
\bibitem{Gatteschi}
D. Gatteschi, A. Caneschi, L. Pardi, and R. Sessoli, Science {\bf 265}, 1054 (1994); 
D. Gatteschi, Adv. Mater. {\bf 6}, 634 (1994); {\it Magnetic Molecular Materials}, Vol. 
198 of NATO {\it Advanced Studies Institute, Series E: Applied Sciences}, edited by D. 
Gatteschi, O. Kahn, J. S. Miller, and F. Palacio (Kluwer Academic, Norwell, MA, 1991).
\bibitem{Friedman}
J. R. Friedman, M. P. Sarachik, J. Tejada, and R. Ziolo, Phys. Rev. Lett. {\bf 76}, 
3830 (1996);
 L. Thomas, F. Lonti, R. Allou, D. Gatteschi, R. Sessoli, and B. Barbara, Nature 
 (London), {\bf 383}, 145 (1996).
\bibitem{V2}
Y. Furukawa, A. Iwai, K. I. Kumagai, and A. Yakubovsky, J. Phys. Soc. Jpn. {\bf
65}, 2393 (1996).
\bibitem{Fe2}
A. Lascialfari, F. Tabak, G. L. Abbati, F. Borsa, M. Corti, and D. Gatteschi,
J. Appl. Phys. {\bf 85}, 4539 (1999).
\bibitem{V3}
A. M{\"u}ller, J. Meyer, H. B{\"o}gge, A. Stammler, and A. Botar, Chem.-Eur. J
{\bf 4}, 1388 (1998).
\bibitem{Nd4}
D. M. Barnhart, D. L. Clark, J. C. Gordon, J. C. Huffman, J. G. Watkin, and
B. D. Zwick, J. Am. Chem. Soc. {\bf 115}, 8461 (1993).
\bibitem{Cr4}
A. Bino, D. C. Johnston, D. P. Goshorn, T. R. Talbert, and E. I. Stiefel,
Science {\bf 241}, 1479 (1988); Y. Furukawa, M. Luban, F. Borsa,
D. C. Johnston, A. V. Mahajan, L. L. Miller, D. Mentrup, J. Schnack, and
A. Bino, Phys. Rev. B {\bf 61}, 8635 (2000).
\bibitem{Tb4}
J. S. Gardner, S. R. Dunsiger, B. D. Gaulin, M. J. P. Gingras, J. E. Greedan,
R. F. Kiefl, M. D. Lumsden, W. A. MacFarlane, N. P. Raju, J. E. Sonier,
I. Swanson, and Z. Tun, Phys. Rev. Lett. {\bf 82}, 1012 (1999).
\bibitem{Fe4}
A. L. Barra, A. Caneschi, A. Cornia, F. Fabrizi de Biani, D. Gatteschi,
C. Sangregorio, R. Sessoli, and L. Solace, J. Am. Chem. Soc. {\bf 121}, 5302 
(1999).
\bibitem{Fe4e}
A. Bouwen, A. Caneschi, D. Gatteschi, E. Goovaerts, D. Schoemaker, L. Sorace,
and M. Stefan, J. Phys. Chem. B {\bf 105}, 2658 (2001).
\bibitem{CaV4O9}
S. Taniguchi, T. Nishikawa, Y. Yasui, Y. Kobayashi, M. Sato, T. Nishioka,
M. Kontani, and K. Sano, J. Phys. Soc. Jpn. {\bf 64}, 2758 (1995); N. Katoh and M. Imada,
J. Phys. Soc. Jpn. {\bf 64}, 4105 (1995).
\bibitem{Fe6}
A. Lascialfari, D. Gatteschi, F. Borsa, and A. Cornia, Phys. Rev. B {\bf 55},
14341 (1997); {\it ibid.} {\bf 56}, 8434 (1997).
\bibitem{Fe8}
D. Gatteschi, A. Lascialfari, and F. Borsa, J. Magn. Magn. Mater. {\bf 185}, 238 (1998).
\bibitem{Fe10}
M.-H. Julien, Z. H. Jang, A. Lascialfari, F. Borsa, M. Horvatic, A. Caneschi,
and D. Gatteschi, Phys. Rev. Lett. {\bf 83}, 227 (1999).
\bibitem{2spin}
D. Mentrup, J. Schnack, and M. Luban, Physica A {\bf 272}, 153 (1999).
\bibitem{Mentrup}
D. Mentrup, H.-J. Schmidt, J. Schnack, and M. Luban, Physica A {\bf 278}, 214
(2000). 
\bibitem{Mueller}
G. M{\"u}ller, Phys. Rev. Lett. {\bf 60}, 2785 (1988).
\bibitem{Landau}
R. W. Gerling and D. P. Landau, Phys. Rev. Lett. {\bf 63}, 812 (1989)
\bibitem{Mueller1}
G. M{\"u}ller, Phys. Rev. Lett. {\bf 63}, 813 (1989).
\bibitem{Bonfim}
O. F. de Alcantara Bonfim and G. Reiter, Phys. Rev. Lett. {\bf 69}, 367 (1992).
\bibitem{Luban}
O. Ciftja, M. Luban, M. Auslender, and J. H. Luscombe, Phys. Rev. B {\bf 60}, 10 122 (1999).
\bibitem{deGennes}
P. G. deGennes, J. Phys. Chem. Solids, {\bf 4}, 223 (1958).
\bibitem{LBC}
M. Luban, T. Bandos, and O. Ciftja (unpublished).
\bibitem{Schroeder}
C. Schr{\"o}der, (Ph. D. Thesis, U. Osnabr{\"u}ck, 1998, unpublished), and
private communications.
\bibitem{LL}
M. Luban and J. H. Luscombe, Am. J. Phys. {\bf 67}, 1161 (1999).
\bibitem{Montroll}
E. W. Montroll, J. Math. and Phys. {\bf 25}, 37 (1946).
\end{references}
\end{document}